\newcommand\farcs{\mbox{$.\!\!^{\prime\prime}$}}

\documentclass[pdflatex,sn-mathphys-num]{sn-jnl}

\usepackage{graphicx}%
\usepackage{multirow}%
\usepackage{amsmath,amssymb,amsfonts}%
\usepackage{amsthm}%
\usepackage{mathrsfs}%
\usepackage[title]{appendix}%
\usepackage{xcolor}%
\usepackage{textcomp}%
\usepackage{manyfoot}%
\usepackage{booktabs}%
\usepackage{algorithm}%
\usepackage{algorithmicx}%
\usepackage{algpseudocode}%
\usepackage{listings}%
\usepackage{caption}
\usepackage{xspace}
\usepackage{CJK}

\newcommand{\hst}{\textit{HST}\xspace}
\newcommand{\euclid}{\textit{Euclid}\xspace}
\newcommand{\jwst}{\textit{JWST}\xspace}
\newcommand{\host}{NGC 6744\xspace}
\newcommand{\sn}{SN\,2024vjm\xspace}
\newcommand{\bpass}{\texttt{BPASS}\xspace}
\newcommand{\ion}[2]{#1\,\textsc{#2}}

\newcommand{\extmat} {\textit{Extended Data}\xspace}
\newcommand{\extmattwo} {Extended Data\xspace}
\newcommand{\suppmat}{\textit{Supplementary Material}\xspace}
\newcommand{\supmattwo}{Supplementary Material\xspace}

\theoremstyle{thmstyleone}%

\theoremstyle{thmstyletwo}%

\theoremstyle{thmstylethree}%

\begin{document}

\title[A Faint Progenitor System for the Faint Supernova 2024vjm]{A Faint Progenitor System for the Faint Supernova 2024vjm}

\author[1*]{\fnm{Erez A.} \sur{Zimmerman}}\email{erez.zimmerman@weizmann.ac.il}
\author[1]{\fnm{Avishay} \sur{Gal-Yam}}
\author[2,3,4]{\fnm{Paul J.} \sur{Groot}}
\author[1]{\fnm{Eran O.} \sur{Ofek}}
\author[5,6]{\fnm{Jan} \sur{van Roestel}}
\author[7]{\fnm{Andrea} \sur{Pastorello}}
\author[8]{\fnm{Stefano} \sur{Valenti}}
\author[8]{\fnm{Aravind P.} \sur{Ravi}}
\author[9,10]{\fnm{Ping} \sur{Chen}}
\author[1,11]{\fnm{Steve} \sur{Schulze}}
\author[12,13,14]{\fnm{Nadejda} \sur{Blagorodnova}}
\author[12,15]{\fnm{Maxime} \sur{Wavasseur}}
\author[12,15]{\fnm{Marco A.} \sur{Gómez-Muñoz}}
\author[12,14,15]{\fnm{Hugo} \sur{Tranin}}
\author[16]{\fnm{Simon} \sur{de Wet}}
\author[16]{\fnm{Giorgos} \sur{Leloudas}}
\author[17]{\fnm{Paul M.} \sur{Vreeswijk}}
\author[11]{\fnm{Lindsey A.} \sur{Kwok}}
\author[18,19]{\fnm{Michaela} \sur{Schwab}}
\author[18]{\fnm{Saurabh W.} \sur{Jha}}
\author[20]{\fnm{Kate} \sur{Maguire}}
\author[21]{\fnm{David J.} \sur{Sand}}
\author[22]{\fnm{Eric} \sur{Stringer}}
\author[22]{\fnm{Thomas} \sur{Kupfer}}
\author[1]{\fnm{Tamar} \sur{Faran}}

\author[23]{\fnm{Joseph P.} \sur{Anderson}}
\author[24]{\fnm{Jennifer} \sur{Andrews}}
\author[25,26]{\fnm{Moira} \sur{Andrews}}
\author[1]{\fnm{Avshalom} \sur{Badash}}
\author[2]{\fnm{Steven} \sur{Bloemen}}
\author[21,27]{\fnm{K. Azalee} \sur{Bostroem}}
\author[28]{\fnm{Ting-Wan} \sur{Chen}}
\author[29]{\fnm{Massimo} \sur{Della Valle}}
\author[30]{\fnm{Georgios} \sur{Dimitriadis}}
\author[31]{\fnm{Yize} \sur{Dong}}
\author[25,32]{\fnm{Joseph R.} \sur{Farah}}
\author[33]{\fnm{James H.} \sur{Gillanders}}
\author[34]{\fnm{Benjamin} \sur{Godson}}
\author[35]{\fnm{Mariusz} \sur{Gromadzki}}
\author[13]{\fnm{Daichi} \sur{Hiramatsu}}
\author[8]{\fnm{Emily} \sur{Hoang}}
\author[25,26]{\fnm{D. Andrew} \sur{Howell}}
\author[36]{\fnm{Daryl} \sur{Janzen}}
\author[37,38]{\fnm{Hanindyo} \sur{Kuncarayakti}}
\author[39]{\fnm{Jiaxuan} \sur{Li}}
\author[34]{\fnm{Joseph D.} \sur{Lyman}}
\author[40]{\fnm{Keiichi} \sur{Maeda}}
\author[34]{\fnm{Mark R.} \sur{Magee}}
\author[25]{\fnm{Curtis} \sur{McCully}}
\author[8]{\fnm{Darshana} \sur{Mehta}}
\author[30]{\fnm{Andrew} \sur{Milligan}}
\author[41]{\fnm{Shane} \sur{Moran}}
\author[25,42]{\fnm{Yuan Qi} \sur{Ni}}
\author[43]{\fnm{David} \sur{O'Neill}}
\author[21]{\fnm{Jeniveve} \sur{Pearson}}
\author[2]{\fnm{Daniëlle L. A.} \sur{Pieterse}}
\author[44]{\fnm{Giuliano} \sur{Pignata}}
\author[7,45]{\fnm{Andrea} \sur{Reguitti}}
\author[46]{\fnm{Daniel E.} \sur{Reichart}}
\author[8]{\fnm{Nicol\'as Meza} \sur{Retamal}}
\author[23,47]{\fnm{Rita P.} \sur{Santos}}
\author[48,29]{\fnm{Simone} \sur{Scaringi}}
\author[49,50]{\fnm{Manisha} \sur{Shrestha}}
\author[33]{\fnm{Shubham} \sur{Srivastav}}
\author[33]{\fnm{Fiorenzo} \sur{Stoppa}}
\author[21]{\fnm{Bhagya} \sur{Subrayan}}
\author[7]{\fnm{Giorgio} \sur{Valerin}}
\author[51]{\fnm{Xiaofeng} \sur{Wang}}
\author[25,32]{\fnm{Kathryn} \sur{Wynn}}
\author[1]{\fnm{Ofer} \sur{Yaron}}
\author[52]{\fnm{Weicheng} \sur{Zang}}

\affil[1]{\orgdiv{Department of Particle Physics and Astrophysics}, \orgname{Weizmann Institute of Science}, \orgaddress{\street{234 Herzl St}, \postcode{7610001}, \city{Rehovot}, \country{Israel}}}
\affil[2]{\orgdiv{Department of Mathematics/IMAPP}, \orgname{Radboud University}, \orgaddress{\street{PO Box 9010}, \postcode{6500 GL}, \city{Nijmegen}, \country{The Netherlands}}}
\affil[3]{\orgdiv{Department of Astronomy and Inter-University Center for Data Intensive Astronomy}, \orgname{University of Cape Town}, \orgaddress{\street{Pivate Bag}, \postcode{X3 7701}, \city{Rondebosch}, \country{South Africa}}}
\affil[4]{\orgname{South African Astronomical Observatory}, \orgaddress{\street{P.O. Box 9, 7935, Observatory}, \country{South Africa}}}
\affil[5]{\orgname{Institute of Science and Technology Austria}, \orgaddress{\street{Am Campus 1}, \city{3400 Klosterneuburg}, \country{Austria}}}
\affil[6]{\orgname{Anton Pannekoek Institute for Astronomy, University of Amsterdam}, \orgaddress{\street{P.O. Box 94249}, \postcode{1090 GE}, \city{Amsterdam}, \country{The Netherlands}}}
\affil[7]{\orgname{INAF – Osservatorio Astronomico di Padova, Vicolo dell’Osservatorio 5}, \orgaddress{\postcode{I-35122}, \city{Padova}, \country{Italy}}}
\affil[8]{\orgdiv{Department of Physics and Astronomy}, \orgname{University of California, Davis}, \orgaddress{\street{1 Shields Avenue}, \postcode{CA 95616-5270}, \city{Davis}, \country{USA}}}
\affil[9]{\orgname{Institute for Advanced Study in Physics, Zhejiang University}, \orgaddress{\city{Hangzhou 310027}, \country{China}}}
\affil[10]{\orgname{Institute for Astronomy, School of Physics, Zhejiang University}, \orgaddress{\city{Hangzhou 310027}, \country{China}}}
\affil[11]{\orgname{Center for Interdisciplinary Exploration and Research in Astrophysics (CIERA)}, \orgaddress{\street{1800 Sherman Ave.}, \postcode{IL 60201}, \city{Evanston}, \country{USA}}}
\affil[12]{\orgname{Institut de Ciències del Cosmos (ICCUB), Universitat de Barcelona (UB), c. Martí i Franquès}, \orgaddress{\street{1}, \postcode{08028}, \city{Barcelona}, \country{Spain}}}
\affil[13]{\orgdiv{Department of Astronomy}, \orgname{University of Florida, Bryant Space Science Center}, \orgaddress{\postcode{FL 32611-2055}, \city{Gainesville}, \country{USA}}}
\affil[14]{\orgname{Institut d’Estudis Espacials de Catalunya (IEEC), c/ Esteve Terradas}, \orgaddress{\street{1, Edifici RDIT, Despatx 212, Campus del Baix Llobregat UPC - Parc Mediterrani de la Tecnologia}, \postcode{08860}, \city{Castelldefels}, \country{Spain}}}
\affil[15]{\orgdiv{Departament de Física Quàntica i Astrofísica (FQA)}, \orgname{Universitat de Barcelona (UB), c. Martí i Franquès}, \orgaddress{\street{1}, \postcode{08028}, \city{Barcelona}, \country{Spain}}}
\affil[16]{\orgname{DTU Space, Technical University of Denmark}, \orgaddress{\street{Building 327, Elektrovej}, \postcode{2800}, \city{Kgs. Lyngby}, \country{Denmark}}}
\affil[17]{\orgdiv{Department of Astrophysics/IMAPP}, \orgname{Radboud University}, \orgaddress{\street{P.O. Box 9010}, \postcode{6500 GL}, \city{Nijmegen}, \country{The Netherlands}}}
\affil[18]{\orgdiv{Department of Physics and Astronomy}, \orgname{Rutgers, The State University of New Jersey}, \orgaddress{\street{136 Frelinghuysen Road}, \postcode{NJ 08854-8019}, \city{Piscataway}, \country{USA}}}
\affil[19]{\orgdiv{Department of Astronomy}, \orgname{University of Virginia}, \orgaddress{\city{Charlottesville VA 22904-4325}, \country{USA}}}
\affil[20]{\orgdiv{School of Physics}, \orgname{Trinity College Dublin, The University of Dublin}, \orgaddress{\city{Dublin 2}, \country{Ireland.}}}
\affil[21]{\orgname{Steward Observatory, University of Arizona}, \orgaddress{\street{933 North Cherry Avenue}, \postcode{AZ 85721-0065}, \city{Tucson}, \country{USA}}}
\affil[22]{\orgname{Hamburger Sternwarte, University of Hamburg}, \orgaddress{\street{Gojenbergsweg 112}, \postcode{21029}, \city{Hamburg}, \country{Germany}}}
\affil[23]{\orgname{European Southern Observatory}, \orgaddress{\street{Alonso de Córdova 3107, Vitacura}, \postcode{Casilla 19001}, \city{Santiago}, \country{Chile}}}
\affil[24]{\orgname{Gemini Observatory}, \orgaddress{\street{670 North A`ohoku Place}, \postcode{HI 96720-2700}, \city{Hilo}, \country{USA}}}
\affil[25]{\orgname{Las Cumbres Observatory}, \orgaddress{\street{6740 Cortona Drive, Suite 102}, \postcode{CA 93117-5575}, \city{Goleta}, \country{USA}}}
\affil[26]{\orgdiv{Department of Physics}, \orgname{University of California, Santa Barbara}, \orgaddress{\postcode{CA 93106-9530}, \city{Santa Barbara}, \country{USA}}}
\affil[27]{\orgaddress{\country{LSSTC Catalyst Fellow}}}
\affil[28]{\orgname{Graduate Institute of Astronomy, National Central University}, \orgaddress{\street{300 Jhongda Road}, \postcode{32001}, \city{Jhongli}, \country{Taiwan}}}
\affil[29]{\orgname{INAF – Osservatorio Astronomico di Capodimonte}, \orgaddress{\city{Napoli}, \country{Italy}}}
\affil[30]{\orgdiv{Department of Physics}, \orgname{Lancaster University, Lancaster}, \orgaddress{\street{LA1}, \city{4YB}, \country{UK}}}
\affil[31]{\orgname{Center for Astrophysics \textbar Harvard \& Smithsonian}, \orgaddress{\street{60 Garden Street}, \postcode{MA 02138-1516}, \city{Cambridge}, \country{USA}}}
\affil[32]{\orgdiv{Department of Physics}, \orgname{University of California}, \orgaddress{\postcode{CA 93106-9530}, \city{Santa Barbara}, \country{USA}}}
\affil[33]{\orgdiv{Astrophysics sub-Department}, \orgname{Department of Physics, University of Oxford, Keble Road, Oxford}, \orgaddress{\street{OX1}, \city{3RH}, \country{UK}}}
\affil[34]{\orgdiv{Department of Physics}, \orgname{University of Warwick, Gibbet Hill Road}, \orgaddress{\city{Coventry CV4 7AL}, \country{UK}}}
\affil[35]{\orgname{Astronomical Observatory, University of Warsaw}, \orgaddress{\street{Al. Ujazdowskie 4}, \postcode{00-478}, \city{Warszawa}, \country{Poland}}}
\affil[36]{\orgdiv{Department of Physics and Engineering Physics}, \orgname{University of Saskatchewan}, \orgaddress{\street{116 Science Place, Saskatoon}, \city{SK S7N 5E2}, \country{Canada}}}
\affil[37]{\orgname{Tuorla Observatory, Department of Physics and Astronomy}, \orgaddress{\street{FI-20014, University of Turku}, \country{Finland}}}
\affil[38]{\orgname{Finnish Centre for Astronomy with ESO (FINCA)}, \orgaddress{\city{FI-20014 University of Turku}, \country{Finland}}}
\affil[39]{\orgname{Department of Astrophysical Sciences}, \orgaddress{\street{4 Ivy Lane, Princeton University}, \postcode{NJ 08540}, \city{Princeton}, \country{USA}}}
\affil[40]{\orgdiv{Department of Astronomy}, \orgname{Kyoto University, Kitashirakawa-Oiwake-cho, Sakyo-ku}, \orgaddress{\city{Kyoto 606-8502}, \country{Japan}}}
\affil[41]{\orgdiv{School of Physics and Astronomy}, \orgname{University of Leicester, University Road}, \orgaddress{\city{Leicester LE1 7RH}, \country{UK}}}
\affil[42]{\orgname{Kavli Institute for Theoretical Physics, University of California, Santa Barbara, 552 University Road}, \orgaddress{\postcode{CA 93106-4030}, \city{Goleta}, \country{USA}}}
\affil[43]{\orgdiv{School of Physics and Astronomy}, \orgname{University of Birmingham, Edgbaston}, \orgaddress{\postcode{B15 2TT}, \city{Birmingham}, \country{UK}}}
\affil[44]{\orgname{Instituto de Alta Investigaci\'on, Universidad de Tarapac\'a}, \orgaddress{\postcode{Casilla 7D}, \city{Arica}, \country{Chile}}}
\affil[45]{\orgname{INAF – Osservatorio Astronomico di Brera}, \orgaddress{\street{Via E. Bianchi 46}, \postcode{I-23807}, \city{Merate (LC)}, \country{Italy}}}
\affil[46]{\orgdiv{Department of Physics and Astronomy}, \orgname{University of North Carolina}, \orgaddress{\street{120 East Cameron Avenue}, \postcode{NC 27599}, \city{Chapel Hill}, \country{USA}}}
\affil[47]{\orgname{CENTRA, Departamento de Física, Instituto Superior Técnico – IST, Universidade de Lisboa – UL}, \orgaddress{\street{Avenida Rovisco Pais 1}, \postcode{1049-001}, \city{Lisboa}, \country{Portugal}}}
\affil[48]{\orgname{Centre for Extragalactic Astronomy}, \orgaddress{\street{Department of Physics, Durham University}, \postcode{DH1}, \city{3LE}, \country{United Kingdom}}}
\affil[49]{\orgdiv{School of Physics and Astronomy}, \orgname{Monash University}, \orgaddress{\city{Clayton}, \country{Australia}}}
\affil[50]{\orgname{OzGrav: The ARC Center of Excellence for Gravitational Wave Discovery}, \orgaddress{\country{Australia}}}
\affil[51]{\orgdiv{Physics Department}, \orgname{Tsinghua University}, \orgaddress{\city{Beijing 100084}, \country{China}}}
\affil[52]{\orgdiv{Department of Astronomy}, \orgname{Westlake University}, \orgaddress{\postcode{Hangzhou 310030}, \city{Zhejiang Province}, \country{China}}}

\abstract{Type Ia Supernovae (SNe Ia) are well known for their role as standardizable cosmological candles. Their uniformity is credited to their single origin as thermonuclear explosions of White dwarf (WD) stars (e.g., \cite{Maoz2014,Jha2019}). Nevertheless, some SNe Ia break this regularity. Prominently, the Iax subclass \cite{Jha2017} are less energetic and remarkably diverse, raising questions about their progenitor systems. While no progenitor system of a normal SN Ia has ever been detected, a luminous blue star was identified in pre-explosion images of the site of the bright SN Iax SN\,2012Z \cite{McCully2014}, suggested to be a helium giant companion star acting as a mass donor to a WD SN progenitor. This is in line with models of weak mass accretion of a WD from a binary companion (e.g., \cite{Fink2014}), producing an explosion that does not fully disrupt the star. However, these models fail to explain the properties of the faintest Type Iax explosions \cite{Kromer2015}, suggesting either they originate from other WD binary systems, or even from massive progenitor stars  \cite{Valenti2009,Moriya2010}. Here, we present the faint SN Iax \sn\,- possibly the faintest supernova observed to date. Using a deep pre-explosion image taken by the recently launched \euclid space mission, we show that its progenitor system must be fainter than the helium giant SN Iax progenitor candidate of SN\,2012Z, as well as that of the luminous red companion or remnant of the faint SN\,2008ha \citep{Foley2014}, and may require a subdwarf helium star as a mass donor. The deep image 
also provides strong arguments against a massive star origin for this faint supernova. Our observations argue that \sn is a WD explosion, but we find that remarkably faint SNe Iax fade more slowly than bright ones, i.e., they evolve in an opposite manner from the famous Phillips relation that makes regular SNe Ia cosmological candles \cite{Phillips1993}.}


\maketitle

\begin{figure}[h!]
    \centering
    \includegraphics[width=\textwidth]{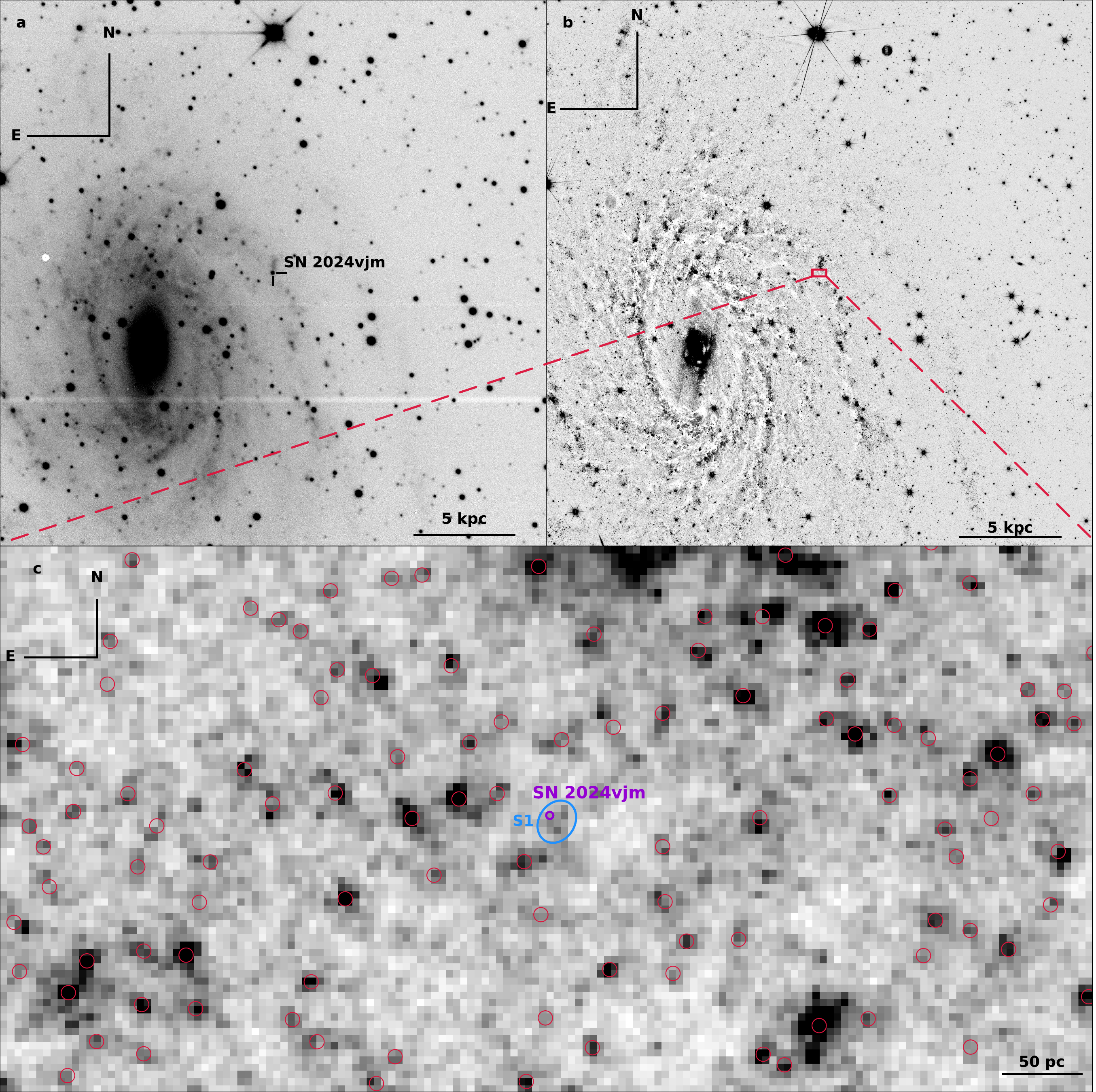}
    \caption{No bright stars at the location of \sn. \textbf{a} post explosion q-band BG image of \sn. \sn is clearly seen at a spiral arm in the north-west of \host (marked by a crosshair)
    \textbf{b}, the pre-explosion \euclid VIS-band image of \host with the nearest $10''$ to the explosion site marked with a red rectangle. \textbf{c}, A zoom in on the 7.5$''$ by 15$''$ around the explosion site. The tip of the large star-forming region to the north of the explosion site is visible in the cutout. The astrometric location of \sn is marked in a 0.1$''$ diameter circle (purple) corresponding to a single \euclid pixel. Sources identified by the \euclid pipeline are marked with 0.2$''$ diameter circles (red). S1, the nearest extended source to the location of \sn, is marked in a cyan ellipse marking its extent. \sn lies just north of S1's centre and is unlikely to be physically associated with the source.}
    \label{fig:Euclid_field}
\end{figure}

\sn was discovered on 2024 September 13 at 23:59:05 UTC ($\rm MJD=60567.00$) by the BlackGEM (BG) Local Transient Survey (LTS;\cite{Groot2024}). The supernova (SN) exploded within a spiral arm in the outskirts of the nearby \host galaxy (distance $9.39\pm0.43$ Mpc, Fig \ref{fig:Euclid_field}b; Methods \ref{sec:environment}), north-west of its centre, at right ascension $\alpha=19^{\rm hr}09^{\rm m}25^{\rm s}.79$ and declination $\delta=-63^{\circ}50^{\prime}01^{\prime\prime}.77$ (J2000).
A classification spectrum was taken on 2024 September 15 at 00:40:56 UTC \citep{Asquini2024}, revealing a plethora of narrow ($v\sim3000\,\rm km\,s
^{-1}$) absorption lines, including low mass elements (LME) such as \ion{C}{II}, \ion{O}{I}, intermediate mass elements (IME) such as \ion{S}{II}, \ion{Si}{II}, \ion{Mg}{II} and \ion{Ca}{II} and iron group elements (IGE), namely \ion{Fe}{II}. The SN was subsequently classified as a Type Iax SN \citep{Srivastav2024} based on its spectral similarity to other faint (reaching peak \textit{B}-band absolute magnitude of $M_{B}>-15$) SNe Iax 2008ha \citep{Valenti2009,Foley2009}, 2010ae \citep{Stritzinger2014}, and 2019gsc \citep{Srivastav2020,Tomasella2020}. Following discovery, we initiated a multiwavelength follow-up campaign to study the SN (Methods \ref{sec:photometry} -- \ref{sec:spectroscopy}), as well as an archival search for pre-explosion imaging of the explosion site. \sn reached peak in B-band brightness on $\rm MJD = 60572.18$, which we use as a reference time (Methods \ref{sec:photometry}).

The host galaxy \host was serendipitously observed by the \euclid space mission \cite{Euclid} as part of an Early Release Observations (ERO) program \citep{euclid_nearby,EROcite} on 2023 October 4 \citep{euclid_pipeline}, a mere 11 months before the discovery of \sn. This presented the opportunity to study the SN explosion site, shown in Figure \ref{fig:Euclid_field}c, in search of its progenitor system. To detect or constrain such progenitor systems, 
we determine the location of the explosion site in the \euclid pre-explosion image to within $10$ milliarcseconds (mas), less than a single \euclid VIS-band ($500$--$900\,\rm nm$) pixel ($0.1''$), using astrometry derived by averaging multiple ground-based seeing-limited images taken by BlackGEM (Methods \ref{sec:progenitor}). Inspecting the \euclid VIS-band image, we find no point source in the SN explosion site pixel. The nearest source detected by the \euclid source identification pipeline in the vicinity of the explosion site is at a separation of $0.12''$, exceeding our astrometric accuracy and corresponding to an offset of at least $5.5\,\rm pc$ in the host itself (Methods \ref{sec:progenitor}). We refer to this source as S1 for clarity. S1 is classified as an extended source, implying that it is inconsistent with the light from a single point source, and originates from diffuse stellar or nebular emission (Methods \ref{sec:progenitor}). S1 is therefore not the SN progenitor, however its light distribution extends into the location of \sn. To place a limit on a possible progenitor at the SN location we measure the brightness of a point source at the SN location, assuming all the light included in the \euclid Point Spread Function (PSF) centred on this location originates from such a putative progenitor rather than S1 (Methods \ref{sec:progenitor}). We find such a point source would have a brightness of $m_{\rm VIS}=26.6\pm0.16$ mag, corresponding to an extinction corrected absolute magnitude of $M_{\rm VIS}=-3.84\pm0.19$ mag. However, since diffuse light from S1 must contribute to the light within the PSF, any progenitor system is likely fainter and this brightness limit is a conservative upper limit.

\begin{figure}[h!]
    \centering
     \includegraphics[width=\linewidth]{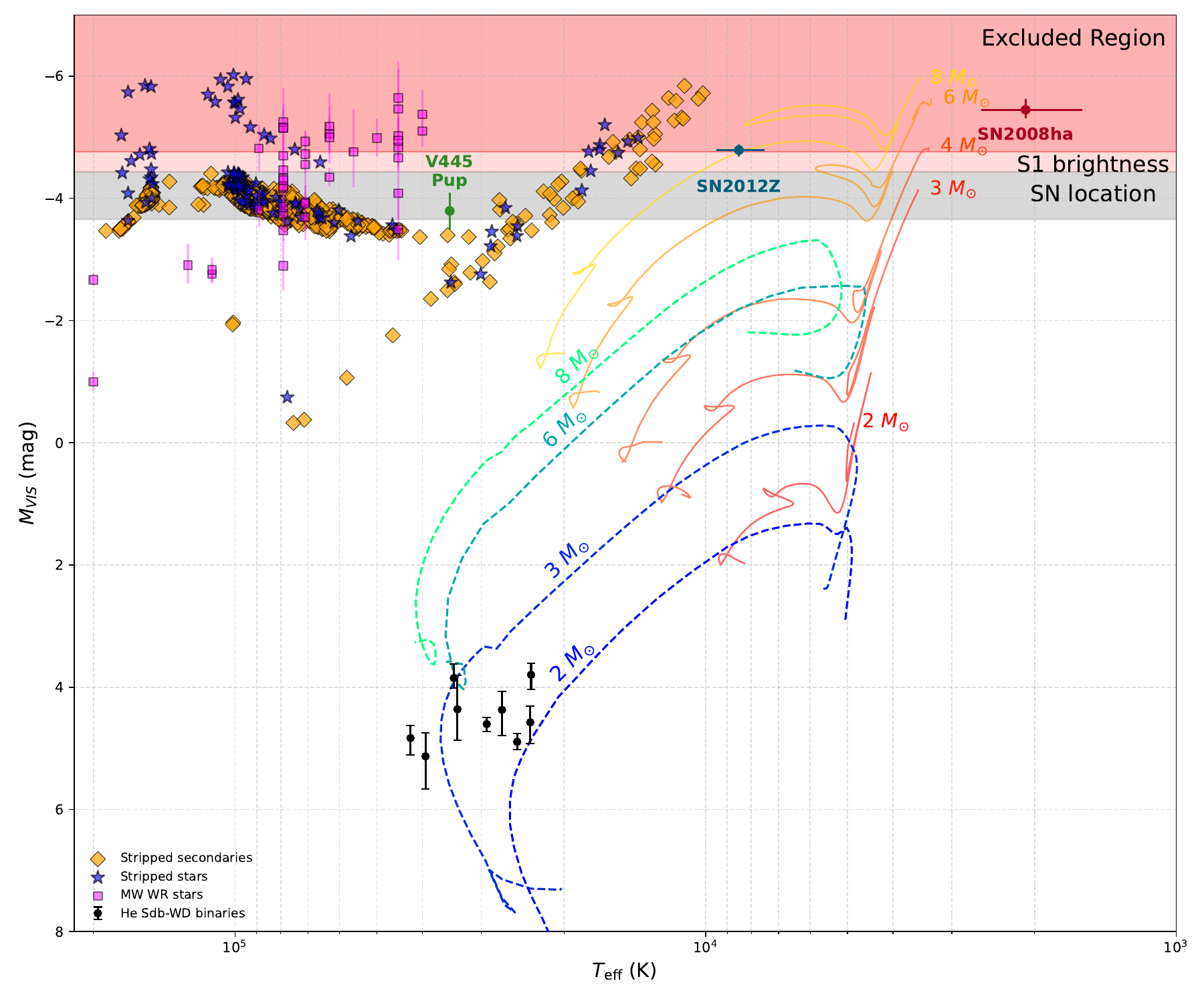}
    \caption{The progenitor system of \sn is fainter than previous Iax progenitor candidates. A Hertzsprung–Russell diagram showing the \textit{Euclid}-VIS band brightness vs temperature. Two sets of \bpass stellar tracks are plotted; single-star H-rich (yellow-red shades) and stripped He tracks (green-blue). The terminal position of massive stripped \bpass tracks is also plotted with blue stars (for primary stars) and orange diamonds (secondary stars). 
    The region occupied by the brightness of S1 is marked in translucent red, and the region occupied by the brightness at the SN location itself is marked in translucent grey. We note that the latter is likely contaminated by S1, and is therefore a strict upper limit of a possible progenitor brightness. We mark the region brighter than both with a darker shade of red. The progenitor system of \sn is fainter than the inferred companion of SN\,2012Z (blue circle), as well as the possible remnant of SN\,2008ha (red circle), and also from that of galactic He Nova V445 Pup (green circle). Known Galactic WR stars \cite{Sander2019} (pink squares) are mostly excluded, though some are slightly fainter than the limit. Low-mass hot subdwarf-WD binary systems (black; Methods \ref{sec:progenitor}) lie well within the permitted brightness range. Such faint systems are possible progenitor systems for faint SNe Iax. Error bars represent $1\sigma$ uncertainties.}
    \label{fig:hr_progenitor_limits}
\end{figure}

From the limit set by the \euclid image (Fig.~\ref{fig:hr_progenitor_limits}), we find that the progenitor system of \sn must have been fainter than previously proposed progenitor systems of SNe Iax (Methods \ref{sec:progenitor}). Specifically, the \sn system was fainter than the luminous blue star coincident with bright Iax SN SN\,2012Z \citep{McCully2014}, suggested to be the He giant companion to the supernova and which may still be visible in deep observations today \cite{Schwab2025}. The progenitor system of \sn is also fainter than the luminous red companion or remnant of the faint SN\,2008ha \citep{Foley2014}, and the Helium (He) star companion of the Galactic He-Nova V445 Pup \citep{Kato2003} as well as the limits placed by the SN\,2014dt progenitor search \cite{Foley2015}. Our work therefore shows, for the first time, that not all SNe Iax progenitor systems contain a luminous source. 

We further compare the \euclid VIS-band brightness limit to theoretical stellar evolution models calculated using the \bpass binary stellar evolution code \cite{Eldridge2017,Stanway2018}. Based on deep observations of the \sn immediate environment using the Multi Unit Spectroscopic Explorer \citep{Bacon2010} (MUSE), we chose sets of models with a solar abundance of metals (Methods \ref{sec:environment}). We use both single star models that simulate the evolution of typical hydrogen-rich stars, as well as binary models leading to stripped stars devoid of hydrogen (H), through Roche-lobe overflow. Comparison to stripped stars is motivated both as massive star progenitors \cite{Moriya2010}, as well as mass-donor companion He-stars to a WD progenitor, similar to the SN\,2012Z progenitor candidate and suggested by theoretical works \cite{Fink2014,Camacho-Neves2023}. Finally, we also compare known Wolf-Rayet (WR) stars \cite{Sander2019} and hot sub-dwarf (SdB)-WD  systems (Methods \ref{sec:progenitor}) observed in our own galaxy to the brightness limit.
The Hertzsprung–Russell (H-R) diagram presented in Figure \ref{fig:hr_progenitor_limits} compares our limits to these systems and the calculated stellar tracks.

Most calculated stellar evolution tracks exceed the luminosity limit and are ruled out. These systems include massive H-rich stars, massive He-star channels (analogous to the SN\,2012Z progenitor candidate), the majority (34 of 48) of WR stars, and the terminal luminosities calculated for most (81\%) massive stripped star progenitor system models. We note that while these massive systems are intrinsically luminous, their spectral energy distributions (SED) peak in the blue, resulting in fainter emission in the redder \euclid-VIS band. Therefore, the effect of dust, while obscuring such sources, would also reprocess emission to the redder bands, mitigating its effect on detectability. Although a subset of massive star systems could evade detection, we find that the absence of other luminous point sources in the vicinity of the explosion site implies a population age inconsistent with the short lifetimes of such progenitors. We therefore consider a massive-star progenitor for \sn unlikely.

Possible progenitor channels for \sn that are left unexcluded are WD systems with faint companions such as low-mass main-sequence or He-star companions, as well as double degenerate systems consisting of just WD stars. 
Specifically, known binary Galactic hot sdB-WD systems are much fainter than our limit. These systems have been proposed to lead to faint thermonuclear explosions \citep{Piersanti2024,Heber2024,Hillman2025,Rajamuthukumar2025} and are therefore interesting candidates for a progenitor system. Regardless, as the permitted companions are not expected to explode as SNe themselves, \sn is likely the result of a thermonuclear explosion of a WD star.

\begin{figure}[h!]
    \centering
     \includegraphics[width=\linewidth]{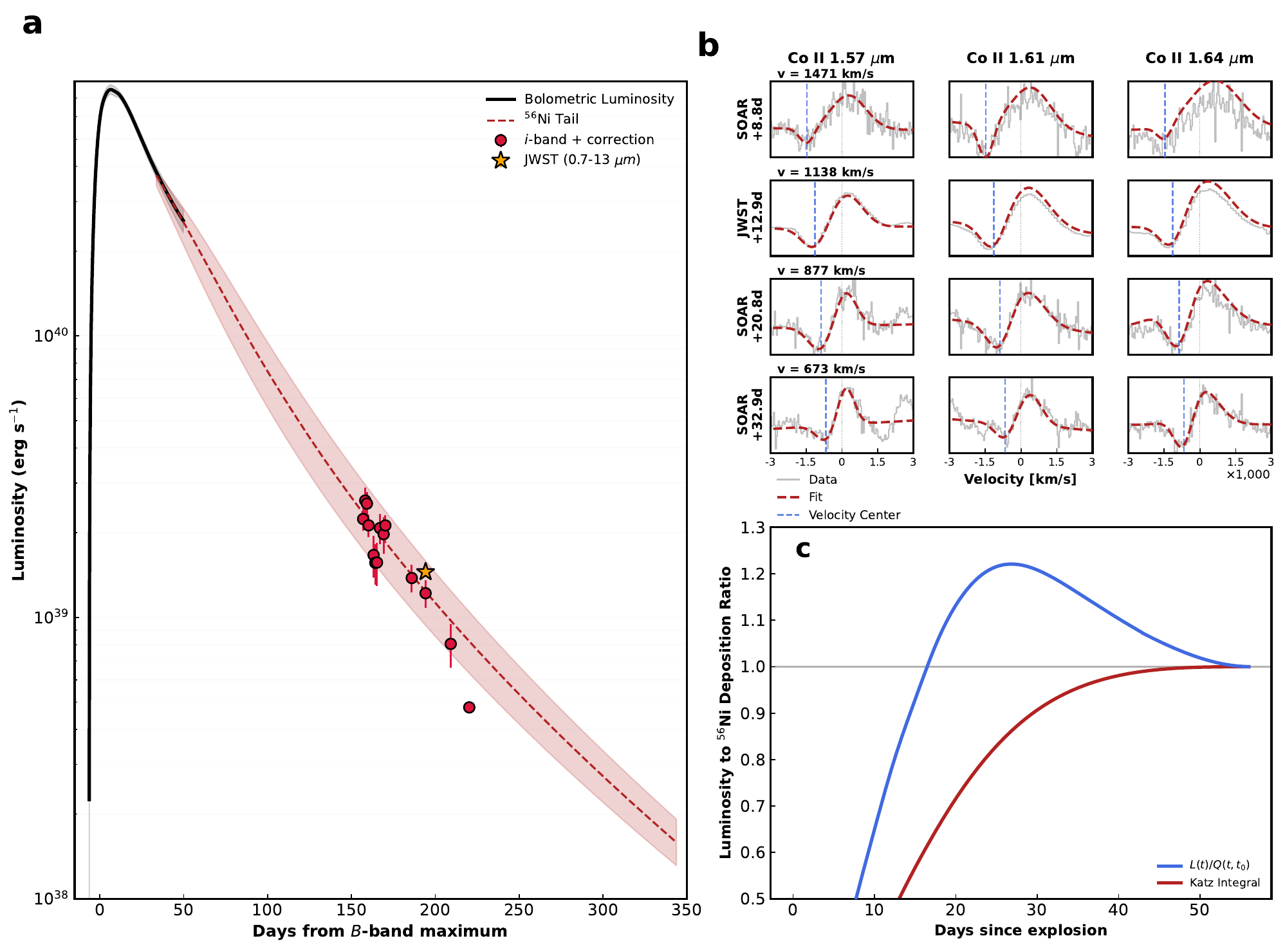}
    \caption{Thermonuclear Diagnostics and Ejecta Evolution of \sn. \textbf{a}. Bolometric light curve and radioactive model. The bolometric luminosity evolution of \sn (black solid line, grey region shows the 1$\sigma$ uncertainties) calculated from multi-band photometry covering the $B$-band through NIR bands. The late light curve is consistent with $^{56}$Ni power (red dashed line). The total flux derived from a $+194$ days \jwst spectrum (yellow star; Methods \ref{sec:power}) agrees with the flux expected from a single $^{56}$Ni decay curve. Late-time $i$-band detections (red circles with 1$\sigma$ errors,), follow the trend of $^{56}$Ni decay,  confirming that the supernova is consistent with a single radioactive deposition curve up to at least $t\approx200$ days. We apply a single bolometric correction to the $i$-band measurements to highlight they follow the $^{56}$Ni trend. As the SED shifts to the red, this correction no longer applies.
    \textbf{b}. NIR \ion{Co}{II} evolution. Sequence of multi-epoch near-infrared spectra (gray steps) centred on the Co~II triplet ($\lambda \lambda$1.57, 1.61, 1.64~$\mu$m). The \ion{Co}{II} features trace $^{56}$Ni decay as $^{56}$Co is its product. A model (red dashed lines) is fit to the Co lines, tracking their development. Blue vertical dashed lines mark the photospheric velocity ($v$) for each epoch, decreasing from $\sim1,500\text{ km s}^{-1}$ to $\sim670\text{ km s}^{-1}$. These exceptionally low velocities show the low-energy explosion failed to rapidly expand the ejecta, producing a slow declining lightcurve.
    \textbf{c}. Katz Integral Energy Deposition Curve. The ratio between the bolometric luminosity ($L$), and the radioactive $^{56}$Ni deposition ($Q$; blue line), as well as the time-integrated ratio of the two (the Katz integral; red line). Both values converge to unity, suggesting no energy source other than $^{56}$Ni decay is needed, with a long $\gamma$-ray escape time ($t_0 \approx 70\text{--}80$ days).}
    \label{fig:Thermo_figure}
\end{figure}

The thermonuclear nature of \sn is further confirmed by the shape of its light curve and spectral features, as shown in Figure \ref{fig:Thermo_figure}. Thermonuclear SNe (i.e. Ia SNe) are powered by the decay of radioactive nickel, the $\rm^{56}Ni$ isotope \cite{Arnett1979,Arnett1982}, through the deposition of energy by $\gamma$-rays and positrons into the ejecta. Using the multi-band lightcurve of \sn, we calculate its bolometric lightcurve (Methods \ref{sec:photometry}) from B-band to the near-infrared (NIR), until \sn fades in B-band beyond detection $\sim 50.7$ days after its B-band peak. We then compare its luminosity, normalised by its time-integration to that of $\rm^{56}Ni$ deposition normalised in the same manner, a method known as the Katz integral \cite{Katz2013} (Methods \ref{sec:power}). This method requires fitting only one parameter - the $\gamma$-ray escape time, $t_{0}$, after which $\gamma$-ray energy deposition becomes inefficient as the ejecta become optically thin. We find the lightcurve strictly adheres to the energy conservation expected from $\rm^{56}Ni$ decay, indicating that $^{56}$Ni is sufficient to explain the lightcurve. Our data do not require any additional power source.

\sn reached a B-band peak magnitude of $M_{B}=-12.7\pm0.1$, when accounting for extinction (Methods \ref{sec:spectroscopy}), making it the faintest SN observed to date, with the possible exception of SN\,2021fcg \cite{Karambelkar2021} for which no colour information is available and dust extinction correction is unreliable. The faintness of \sn indicates that only a small amount of $\rm^{56}Ni$ was synthesised by the explosion. To measure this nickel mass, we further fit a $\rm^{56}Ni$ deposition curve directly to the late-time bolometric curve (the nickel tail; Methods \ref{sec:power}). We find good agreement between the best-fit $t_{0}$ value determined using this method and the values previously inferred from the Katz integral, and measure a $^{56}$Ni mass of $3.6\times10^{-3}\rm\,M_{\odot}$. This value is similar to that derived for other faint SNe Iax \cite{Foley2009,Stritzinger2014,Karambelkar2021}, and is two orders of magnitude lower than regular SN Ia $\rm^{56}Ni$ yields \cite{Stritzinger2006}.

While the low nickel yield is expected in a low-energy explosion, the bolometric analysis reveals an unexpectedly long timescale for $\gamma$-ray trapping in the SN ejecta. Both the tail fit and the Katz integral method yield a $\gamma$-ray escape time of $70$--$80$ days (Methods \ref{sec:power}), meaning the ejecta remained optically thick to $\gamma$-rays post-explosion. This value is significantly longer than for regular type Ia SNe where the $\gamma$-ray escape time is typically $t_{0}\approx30-45$ days \citep{Wygoda2019,Sharon2020} and larger $^{56}$Ni yields are correlated with longer $\gamma$-ray escape times.

\begin{figure}
    \centering
    \includegraphics[width=\linewidth]{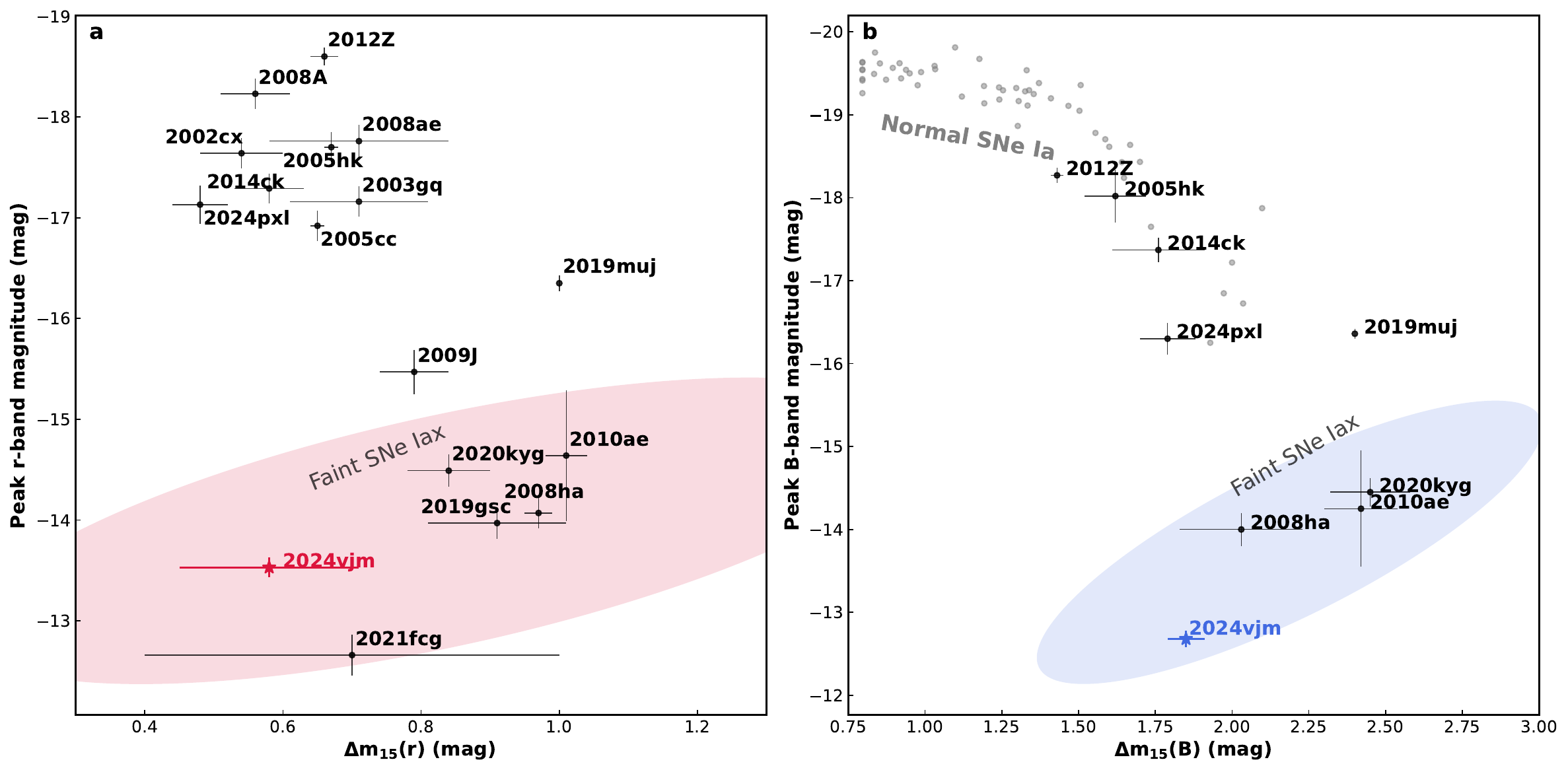}
    \caption{Peak absolute magnitude versus 15-day magnitude decline for \sn. 
            \textbf{a.} Absolute peak $r$-band magnitude ($M_r$) versus the 15-day decline rate ($\Delta M_{15}(r)$). 
            \textbf{b.} Absolute peak $B$-band magnitude ($M_B$) versus $\Delta M_{15}(B)$. 
            In both panels, \sn is denoted by a coloured star, while comparison Type SNe Iax are shown as black circles. The shaded regions highlight other particularly faint SNe Iax. Panel \textbf{b} also shows regular Type Ia supernovae from the Carnegie Supernova Project \cite{Krisciunas2017} (gray points), that follow the famous Phillips relation, whereby fainter events fade faster than bright ones. \sn however fades slower than brighter Iax events, 
            due to its low velocity ejecta. The physics responsible for the decline trend of the Phillips relation holds, perhaps, for bright SNe Iax, but not for the faintest events like \sn. Error bars represent $1\sigma$ uncertainties.}
            \label{fig:phillips}
\end{figure}

As shown in Figure \ref{fig:phillips}, \sn breaks this trend as its low $^{56}$Ni ejecta remained optically thick to $\gamma$-rays for a long duration after explosion. Observationally, this causes the lightcurve of \sn to fade more slowly than its peak brightness would suggest (Methods \ref{sec:photometry}). The initial decline rate of Type Ia SNe is parameterised by $\Delta M_{15}$ -- the drop in magnitudes 15 days after peak. 
Regular Type Ia SNe follow the Phillips relation \cite{Phillips1993} whereby fainter events fade more quickly (larger $\Delta M_{15}$). While SNe Iax do not follow this exact relation due to their heterogeneous nature, bright SNe Iax still typically follow a trend where fainter SNe fade more quickly \cite{Singh2023}. However, \sn reverses this behavior entirely. Compared to other well-observed faint SNe Iax, it fades significantly more slowly (smaller $\Delta M_{15}$ in all optical bands; Methods \ref{sec:photometry}). As \sn is fainter than other SNe Iax, its slow decline robustly confirms the hypothesis by ref. \cite{Singh2023} that faint Iax SNe follow a trend opposite to the Phillips relation.

We identify the cause for the ``faint-yet-slow" evolution of \sn as the inability of the low-energy explosion to rapidly expand its ejecta (Methods \ref{sec:power}), keeping the ejecta optically thick to $\gamma$-rays. The $\gamma$-ray escape time of Type Ia SNe ejecta is governed by the average ejecta column density ($\Sigma$) \cite{Guttman2024}, which scales as $\Sigma\propto M_{ej}\, v^{-2}$, where $M_{ej}$ is the ejecta mass and $v$ is the ejecta expansion velocity. The NIR and optical spectra of \sn reveal that most features expand in an exceptionally low velocity (Methods \ref{sec:spectroscopy}), ranging from $v\approx3000\,\rm km\,s^{-1}$ measured in the earliest spectroscopic epoch to $v<1000\,\rm km s^{-1}$ 20 days post B-band peak brightness. Of particular interest are the distinct \ion{Co}{II} features in the NIR spectra of \sn, since $\rm^{56}Co$ is the product of $\rm^{56}Ni$. The slow expansion velocity of these \ion{Co}{II} features therefore demonstrates that the radioactive products themselves are moving slowly. This slow evolution is also consistent with that of \ion{Ni}{II} in the mid-infrared spectrum of \sn taken with the \textit{James Webb Space Telescope} (\jwst) \cite{Kwok2025}.

While other faint SNe Iax have also shown low-velocity expansion, the low ejecta mass characterizing faint SNe Iax \cite{Foley2013} can compensate for the effect of their slow ejecta expansion on $\gamma$-ray trapping. With a slow expansion velocity and a large $t_{0}$ we find that the ejecta mass of \sn must be of order $M_{ej}\approx0.1\,\rm M_{\odot}$. To confirm this by an independent measurement, we apply a scaling relation between the luminosity rise time, sensitive to the photon diffusion time through the ejecta, and the ejecta mass \cite{Foley2009}. The measured rise-time of $\sim 11$ days indicates an ejecta mass of $M_{ej}\approx0.1\,\rm M_{\odot}$, confirming the low ejecta mass does not compensate for the low expansion velocity in determining the slow evolution of \sn. 

Our constraints on the ejecta mass provide a strong confirmation to previous suggestions regarding the existence of surviving bound remnants left behind Iax SN explosions due to their ejecta-mass budget \cite{Foley2013}, as well as very late-time luminosity excess \cite{Kawabata2021,Maeda2022}. Our evidence for a WD progenitor system confirms that \sn must have left behind a gravitationally bound remnant, as even the lowest mass WD systems \cite{Kilic2007} have more mass than the estimated ejecta mass produced by \sn. Despite this, we find that the late-time lightcurve of \sn shows no additional excess luminosity beyond that expected from a single thermonuclear explosion, and follows its expected decay rate (Methods \ref{sec:power}). This suggests that, in contrast to brighter members of the class, where a surviving remnant may contribute additional luminosity via secondary ejecta or disruption of the remnant \cite{Kawabata2021}, the remnant of \sn did not contribute significant energy to its light curve. We further find the optical late-time lightcurve to be very red at $\sim 200$ days post peak (Methods \ref{sec:power}), possibly due to dust formation, which Iax SNe have been suggested to efficiently produce \cite{Kumar2025}. The SED shifts even more to the red in later epochs ($\sim 350$ days), reminiscent of dust accumulation. Deep infrared studies of \sn with space telescopes such as \jwst would better probe dust formation and possible later signatures from the bound remnant.

Faint SNe Iax occupy the very faint end of the SN population as a whole. Our work provides a strong indication that the progenitor systems of these faintest known SNe are themselves also faint and likely composed of low-mass binary systems. The advent of deep wide-field surveys such as BlackGEM and, in particular, the Rubin Observatory’s Legacy Survey of Space and Time (LSST) \cite{LSST2019}, could put this suggestion to a strong test, providing deep multicolour observations of the sites of similar future events. When combined with deep, wide-field space surveys such as \euclid and the future Nancy Grace Roman Space Telescope \cite{Schlieder2024}, the progenitor systems of such events, and those of brighter SNe, will be directly constrained. \sn provides unprecedented high-quality observations of an event within a previously unexplored regime of very faint Type Iax explosions, placing strict new constraints on the allowed progenitor systems and demonstrating that such objects follow different physical correlations compared to both brighter SNe Iax and normal SNe Ia.  

\newpage

\section{Figures}\label{sec6}
\newpage

\newpage

\newpage

\clearpage

\section{Methods}\label{methods}
\subsection{The environment of \sn}\label{sec:environment}
\subsubsection{The host galaxy}

We adopt a distance of $9.39\pm0.43$ Mpc to \host, derived by tip of the red giant branch analysis of the PHANGS-HST survey \cite{Anand2021}. This corresponds to a distance modulus of $29.86\pm0.1$.
For the redshift, we adopt the value of $z=0.0028$ reported by NED. This value matches the SN spectral features and the host's interstellar \ion{Na}{I} D doublet observed in the SN spectra.

\subsubsection{Extinction}
We adopt the Galactic foreground extinction from NASA Extragalactic Database (NED) \cite{ned} of $E(B-V)=0.043$ mag based on ref. \cite{Schlegel1998} and recalibrated by ref. \cite{Schlafly2011}.
To measure the host galaxy extinction, we use the relations from \cite{poznanski2012} between the sum equivalent width (EW) of the \ion{Na}{I}~D doublet and $E(B-V)$ reddening. To measure the EW of the lines  we use the --2.1 days X-shooter spectrum being the highest resolution spectrum in our dataset. We first remove the continuum by fitting a 3rd degree polynomial to the area around the \ion{Na}{I}~D doublet. We then directly measure the EW of each line in the spectrum by integrating the area up to $2\,{\rm \AA}$ around the line. We measure $EW(D1)=0.47\, {\rm\AA}$ and $EW(D2)=0.52\,{\rm \AA}$ corresponding to an extinction value of $E(B-V) = 0.21$ mag.

\subsubsection{Immediate environment and metallicity}
To study the immediate environment of \sn, we obtained a single epoch of observations with the Multi Unit Spectroscopic Explorer (MUSE) \cite{Bacon2010} instrument mounted on the Very Large Telescope (VLT) in the ESO Paranal observatory through a director’s discretionary time (DDT) program 114.28GZ (PI Zimmerman). The observations were conducted using the instrument's Narrow Field Mode (NFM) utilizing Adaptive Optics (AO) \cite{Arsenault2008,2012SPIE.8447E..37S}. This mode allows observation with a spaxel resolution as small as 25 mas, meaning that at the distance of \host each spaxel represents just $1$ pc upon the host. Since no bright star exists within the narrow $7.5''$ field of view, we used \sn itself as a tip-tilt star for the AO guiding. Further details on the reduction process are described in the supplementary material section \ref{sec:MUSE_red}.

\sn exploded within a spiral arm in the outskirts of \host north-west of its centre. The $\rm H\alpha$ MUSE slice, presented in Extended Data Figure \ref{fig:MUSE_Ha}, shows that \sn lies approximately 90 pc south of the edge of a large star formation region, suggesting recent star formation in the vicinity of the SN site. A second, more diffuse star-forming region lies approximately 165 pc southwest of \sn, as well as diffuse star-formation directly west of \sn. This suggests that while the explosion site does not lie in a region with ongoing star formation, it is surrounded by star-forming regions and therefore a young stellar population. This is in line with the explosion sites of other SNe Iax, which are often found near star formation in their host galaxies \citep{Lyman2018}.

We further analyze the strong emission lines in the star-forming region to the north. To gather enough signal for this analysis, we sum the contribution of a 70 spaxel circle capturing the star-forming region to the north of \sn. We find that many emission lines are well detected in the summed spectrum including $H\alpha$, $H\beta$, [\ion{O}{III}] $\lambda5008$, [\ion{N}{II}] $\lambda \lambda\,6548,6583$ and [\ion{Si}{II}] $\lambda\lambda\, 6716, 6730$.
We then measure the flux of each line by fitting a Gaussian to each emission line and integrating the total emission as set by the best-fit Gaussian. We present the result of this measurement in \suppmat Table \ref{tab:host_emission_lines}. We attempted to make the same measurement around the more distant south-west star-forming region; however, the signal for this region was low, and many lines were not well detected.

We use the measured fluxes to assess the metallicity of the star-forming region using the O3N2 metallicity indicator, and adopting the calibration of \cite{Curti2017}. We find the 
oxygen abundance to be $12+\rm log(O/H)=8.73\pm0.09 \,\rm dex$.
Compared to the solar composition \cite{Asplund2009}, this indicates a metallicity of $\frac{\rm[Fe/H]}{\rm[Fe/H]_{\odot}}=1.09^{+0.25}_{-0.20}$. We therefore conclude the immediate environment of \sn is consistent with a solar abundance of metals.

SNe Iax have been found in solar abundance environments in the past \citep{Yamanaka2015,Magee2017}. In particular, they have been shown to follow a similar metallicity distribution as core-collapse SNe \citep{Lyman2018}.
However, the metallicity measurement around \sn is in contradiction with the hypothesis raised by ref. \cite{Srivastav2020}, that faint SNe Iax explode in more metal-poor environments. This was based on their measurement of sub-solar metallicity at the host of SN\,2019gsc and the slightly sub-solar environment measurements of SN\,2008ha \citep{Foley2009} and SN\,2010ae \citep{Stritzinger2014}. Regardless, all faint Type Iax explosion sites are found near star-forming regions and hence younger stellar populations.

\subsubsection{Radio}
We obtained a single epoch of radio continuum follow-up observations with the MeerKAT radio telescope \citep{Jonas2016} in the L-band (900--1670~MHz) and S4-band (2625-3500~MHz) through DDT proposal DDT-20240920-SD-01 (PI de Wet). The total integration time on source was one hour each for the L and S4-band observations, with J1939-6342 used as the flux, bandpass and gain calibrator. The mid-times of the S4-band and L-band observations were 18:26 and 20:17 UT on 2024 September 21. Inspecting the SARAO Science Data Processor (SDP) images, we do not detect a radio source at the position of SN 2024vjm. The median RMS noise was 17 and 11~$\mu$Jy in the L and S-band images from which we derive $3\sigma$ upper limits of 51 and 33~$\mu$Jy, respectively, on any radio emission. This corresponds to luminosity limits of $4.9$ and $7.6\times10^{33}\,\rm erg\,s^{-1}$ in the L and S-band respectively.

\subsection{Search for progenitor system}\label{sec:progenitor}
We conducted an archival data search for deep pre-explosion images of \sn's location. While the host \host was observed in the past by \textit{HST}, we found the SN location within the galaxy has not been previously observed. However, \host was observed by the \euclid space mission \cite{Euclid} as part of their Early Release Observations (ERO) program \cite{euclid_nearby,EROcite}. Owing to the mission's large field of view, the entire galaxy was visible, including the SN location.

\subsubsection{Precision Astrometry}
It is common to achieve precise astrometry using another space-based or adaptive optics imaging. However, here we present a careful astrometric analysis of the seeing-limited BlackGEM (BG) images.
To improve upon the astrometric solution derived by the standard BG pipeline for single images, we derived a joint astrometric solution based on 234 $q$-band images and 54 $i$-band images. The solution was calculated using the \textit{AstroPack/MAATv2} tools\cite{Ofek2014_MAAT, Ofek2019_Astrometry_Code,Soumagnac+Ofek2018_catsHTM} developed for the Large Array Survey Telescope (LAST) \cite{Ofek+2023PASP_LAST_Overview, BenAmi+2023PASP_LAST_Science} and the ULTRASAT space mission \cite{Shvartzvald+2023PASP_ULTRASAT_Overview}.
The astrometry was conducted with respect to the GAIA-DR3 catalog \cite{GAIA+2016_GAIA_mission,GAIA+2022yCat_GAIA_DR3_MainSourcesCatalog}, considering the sources' proper motion and parallax.

For each BG image, a section of $1001\times1001$ pixels was cropped around the approximate SN position. Next, we searched for sources, measured their positions via PSF-fitting, and solved the astrometry for each image. The astrometry based on Gaia DR3 was solved for each field, including distortions represented by third-order polynomials. This is enough to represent differential atmospheric refraction to better than 1\,milliarcsecond (mas). The typical asymptotic rms (i.e., the terms of the astrometric solution at the bright end of $\lesssim18$\,mag) was about 15\,mas. Next, we matched the sources over all images by coordinates. For each source, we calculated its median position and its standard deviation (std) across all images obtained with a specific filter.
The median astrometric position of identified Gaia sources in $i$-band is presented in Extended Data Figure \ref{fig:astrometry_i_band}, showing the improvement in astrometric noise by binning multiple images. 
At the bright end, the typical rms per image is about 10\,mas, and can reach at least 3\,mas by binning a large number of observations.
However, this method does not consider chromatic aberrations (in the atmosphere or telescope). Therefore, we adopted the median $i$-band position (which is less affected by chromatic effects) as the final SN position. The difference between the median $i$-band and $q$-band position of the SN was less than 9\, mas (in each axis), while the robust std of the $i$-band measurements was about 20\,mas (in each axis).
However, since this is larger than the possible systematic effect (e.g., chromatic), we adopted a 10\,mas error in each axis. The adopted positions are:
$\alpha=19^{\rm hr}09^{\rm m}25^{\rm s}.7876$ and $\delta=-63^{\circ}50^{\prime}01^{\prime\prime}.7718$

\subsubsection{Brightness limits from \euclid}
To search for a progenitor candidate in the \euclid data we analyze the VIS-band image, since it has a smaller pixel size of just $0.01''$ instead of $0.02''$ in the \textit{Y},\textit{J} and \textit{H} bands and typically goes $1.5$ mag deeper \cite{euclid_pipeline}.
The astrometric accuracy of \euclid ERO has been shown to be better than $10$ mas \citep{euclid_pipeline} in the VIS band, thus the SN location can be associated with a single pixel within the \euclid VIS image.

\euclid employs multiple source detection algorithms to identify astrophysical objects in its imaging data \citep{euclid_pipeline}. S1, along with other sources marked in Figure \ref{fig:Euclid_field}, were identified by the {\tt SExtractor} \cite{Bertin_sextractor} catalog distributed in the \euclid ERO. We note that other elongated sources are not detected in the catalog and that S1 itself is flagged as an elongated source with a {\tt CLASS\_STAR} value of $0.09$. This suggests that S1 is the result of the congregation of multiple stellar systems or diffuse star formation too faint to be detected in our MUSE data. The total brightness of S1 as determined by a Kron-like automated aperture magnitude ({\tt MAG\_AUTO}) is $m_{VIS}=25.83\pm0.13$ mag. 
Since \sn lies within the extent of S1, its location is likely contaminated by its light. Therefore, to obtain an upper limit to the brightness of a possible progenitor system we perform a forced PSF-photometry measurement at the SN location. We use the PSF shape distributed as part of the ERO, the \euclid pipeline weights file for the background error-estimation, and the gain and zeropoint values from the image header, consistent with the values for \host in the \euclid ERO pipeline paper \cite{euclid_pipeline}. We perform the PSF photometry using the {\tt PSFPhotometry} function in the {\tt Photutils Python} package \cite{larry_bradley_2025_14889440}. To verify our custom pipeline we compare our measurement to that of all point sources ({\tt CLASS\_STAR $>0.8$}) within $20''$ of the SN location in the \euclid catalogue. We find a mean offset of $0.06$ mag between our measurement and that of the catalogue {\tt MAG\_PSF} values, confirming its accuracy. We then measure a brightness of $m_{\rm VIS} = 26.58\pm0.16$ mag at the SN location, which can be used as a strict upper limit to any progenitor system, if it were not contaminated by S1.

We compared the VIS magnitude limits to theoretical \bpass models of binary stellar evolution. Specifically, we use the v2.2.1 set of stellar tracks \cite{Eldridge2017,Stanway2018} with a solar abundance of metals ($Z=0.014$). To probe hot sdB progenitor companion star channels, we chose to plot the full tracks of a subset of low ($<8\,\rm M_{\odot}$) mass stripped stars according to the following criteria: (a) The stars are secondary stars, (b) The primary has turned into a $\sim 1\,\rm M_{\odot}$ WD, (c) They have no surface hydrogen. For the rest of the stripped stars, we only plot their end-state to probe them as direct progenitors, rather than binary companions.
We calculate the \euclid-VIS band magnitude for each track by estimating the SED from the calculated standard filter brightness provided in the \bpass code. We then calculate the \euclid-VIS absolute magnitude by running synthetic photometry on the SED using the \euclid-VIS band transmission function. We also plot the full evolutionary tracks for a subset of single-star models ($>8\rm \,M_{\odot}$) in Figure \ref{fig:hr_progenitor_limits}. 

For the SN\,2012Z \cite{Stritzinger2014} and SN\,2008ha \cite{Foley2014} systems, we calculate the perceived \euclid-VIS magnitudes by fitting a blackbody to the corresponding \hst detections and deriving their synthetic photometry in the \euclid VIS-band. For the V445 Pup companion \citep{Kato2003,Woudt2009}, we assume a blackbody at a temperature of $35$ kK. Assuming a colder temperature pushes the companion further into the excluded region, as the SED becomes redder.
Figure \ref{fig:hr_progenitor_limits} also shows WR stars from ref. \cite{Sander2019}, for which we apply a V-band to VIS-band correction, which corrects the underlying blackbody SED instead. This is to avoid excluding additional flux from V-band emission lines that can contribute substantial flux in WR stars. Since this method does not take into account corrections from red emission lines, it may be underestimating the stars' brightness.
We also plot several known Milky Way subdwarf(SD)-WD binary systems: CD -30 11223\citep{Geier2013}, PTF J0823 \citep{Kupfer2017b}, OW J0741 \citep{Kupfer2017a}, ZTF J2130 \citep{Kupfer2020}, ZTF J2055 \citep{Kupfer2020b}, HD~265435 \citep{Pelisoli2021}, PTF J2238 \citep{Kupfer2022}, LAMOST J1710 \citep{Yang2025} and ZTF J0007 \citep{Stringer2025}, for which we also calculate synthetic VIS-band photometry assuming their SED is a blackbody. These stars are located next to corresponding \bpass binary stellar tracks. Specifically, Both CD -30 11223 and PTF J2238 were found in young stellar population, similar to the environments of type Iax SNe.

\subsection{Photometry}\label{sec:photometry}
\subsubsection{Light curve properties}
Following discovery, we initiated a photometric campaign to obtain a pan-chromatic lightcurve. We present the bolometric light curve of \sn in Extended Data Figure \ref{fig:lightcurve}.
To measure the light curve properties of \sn, we interpolate our extinction-corrected multi-band light curve using a custom Gaussian process (GP) fitter based on the \texttt{sklearn} \citep{scikit-learn} \texttt{Python} Package. We present further details on this interpolation in the supplementary material section \ref{app:lc}. This method resulted in a stable interpolation, which phenomenologically follows the band curves, and allows us to measure the time of maximum light, maximum brightness, and decline rate ($\Delta M_{15}$) for each band. We omit the $u$-band from this calculation as it is not well sampled, due to the low $u$-band brightness of \sn. We present the measured lightcurve properties in Extended Data Table \ref{tab:lc_prop}.

Using the interpolated multi-band light curve, we measure the bolometric luminosity of \sn by directly integrating it, achieving an almost complete bolometric curve (as \sn is not blue and we cover \textit{B}-band to \textit{K}-band). Figure \ref{fig:Thermo_figure} panel a shows the bolometric light curve.

With a peak magnitude of $M_{B} = -12.68 \pm 0.1$, \sn is arguably the faintest supernova observed to date. The Type Iax SN\,2021fcg \cite{Karambelkar2021} may be fainter; however, that event was not well-observed and suffers from high extinction uncertainty. \sn occupies a luminosity regime similar to Gap Transients  \cite{Pastorello2019}. Notably, ILOT M85 OT2006-1 \cite{Kulkarni2007}, commonly considered a stellar merger, has also been proposed as a core-collapse SN \cite{Pastorello2007}. Such a scenario would make M85 OT2006-1 a fainter candidate than \sn. M85 OT2006-1 is, however, physically different from faint SNe Iax, occurring in an S0 galaxy (rather than the star-forming hosts of SNe Iax) and showing a plateau light curve inconsistent with a thermonuclear explosion.

\subsubsection{Explosion time estimate}
To estimate the explosion time, we fit a broken power law $f(t) = a\times(t-t_{exp})^n$ to the early $q$-band light curve. We chose the $q$-band since it has the earliest data points, as well as a single pre-detection measurement retrieved by the BG forced photometry pipeline. We use both the $q$-band data from BG and MeerLICHT. Since MeerLICHT has observed \sn with a higher cadence, we bin the data points to improve the measurement. Our fit returns an explosion time of $\rm MJD=60565.58 \pm 0.31$, 6.6 days before \textit{B}-band maximum with a power law index of $n=0.91\pm0.13$. This result is consistent with those of ref. \cite{Magee2025}, which are based on the GOTO lightcurve.

\subsubsection{Slow photometric decline}
Figure \ref{fig:phillips} shows the peak brightness of \sn and other SNe Iax \cite{Foley2013, Stritzinger2014, Stritzinger2015, Tomasella2016, Srivastav2020, Karambelkar2021, Barna2021, Singh2023, Singh2025} in absolute peak magnitudes vs its $\Delta M_{15}$ value in r-band and B-band. When compared to other faint SNe Iax we find that \sn declines quite slowly 15 days post peak. This effect would be in line with the unusually high $t_{0}$ parameter we measure for \sn as a slow decline implies a more opaque ejecta.

With the addition of \sn, we find two distinct groups of SNe Iax in the absolute magnitude vs decline rate graph. Namely, while bright SNe Iax ($M_{r}\lesssim-15$) decline more slowly with brightness, the faint group of SNe Iax ($M_{r}\gtrsim-15$) declines faster the brighter the SN is. This was proposed before \cite{Singh2023,Magee2016}, though with a marginal statistical significance. With \sn being possibly the faintest of this group, it probes a new part of this parameter space, strengthening this correlation (with a linear Pearson correlation coefficient of $-0.81$ and a \textit{p}-value of $0.052$, i.e. a $\sim 2\sigma$ correlation).

\subsection{Spectroscopy}\label{sec:spectroscopy}
After classification, we obtained a sequence of 20 additional optical spectra and three near-infrared (NIR) spectra. Optical spectra are presented in Extended Data Figure \ref{fig:spec} and the NIR spectra are presented in Extended Data Figure \ref{fig:NIR_spec}. 

\subsubsection{Spectral Evolution}
Our spectral series ranges from --4.22 days to 70.8 days before and after the \textit{B}-band maximum. \sn showed spectroscopic similarity to other faint SNe Iax, sharing most, if not all, spectral features, albeit with lower expansion velocities. A comparison between the spectra of \sn and two other faint SNe Iax is shown in Extended Data Figure \ref{fig:spec_comparison}.

The first classification spectrum shows a plethora of narrow absorption lines. Notably, we identify strong low mass elements (LME) and IME such as \ion{C}{II}, \ion{O}{I}, \ion{Mg}{II}, \ion{Si}{II}, \ion{S}{II} and strong \ion{Ca}{II} NIR triplet with P Cygni profiles. From the IGE, we identify relatively few \ion{Fe}{II} transitions, while we find no strong evidence for more highly ionized IGEs, such as \ion{Fe}{III}. In the NIR, this is even more striking before peak, with only two features identified in the pre-peak X-shooter spectrum being \ion{Mg}{II} $\lambda 10927$ in $J$-band and a weak \ion{Fe}{II} $\lambda 16907$ emission line in the H-band. Weak signs of \ion{Co}{II} may be present as well; however, it is hard to positively identify them.

After peak, we find more IGE lines developing, as the number of overall lines increases significantly. We find many of those lines to be consistent with \ion{Fe}{II}, \ion{Cr}{II}, \ion{Co}{II} and \ion{Ti}{II}. A comprehensive study of a +12 days \jwst IR spectrum, including comprehensive line identification was conducted by ref. \cite{Kwok2025}.

In the NIR, the appearance of \ion{Co}{II} lines after peak is striking, as they dominate the H and K bands with distinct P Cygni profiles. We also identify other IGE developing in the J band (for detailed NIR line identification, refer to ref. \cite{Kwok2025}). The least blended set of \ion{Co}{II} transitions (not significantly overlapping with Fe lines) are the $15759,16064$ and $16361\,\rm \AA$ transitions. These lines are usually blended in regular Type Ia SNe, though the low velocity regime of fainter SNe Iax makes these features separate (as seen in the NIR spectra of e.g. SN\,2024pxl, \cite{Kwok2025,Hoogendam2025,Singh2025}, 
SN\,2019muj, \cite{Barna2021}, and SN\,2010ae, \cite{Stritzinger2014}).
The appearance of \ion{Co}{II} lines only after peak is intriguing. Our first NIR spectrum was taken about four days after the explosion; therefore, not even half of the Co abundance has been synthesised at the time, as the half-life of $^{56}$Ni is $\sim 6$ days. This later appearance is also consistent with the other faint SNe Iax \cite{Stritzinger2014}. Although forbidden \ion{Co}{III} is detected in the NIR \jwst spectrum\cite{Kwok2025}, the lack of permitted \ion{Fe}{III} lines in the early spectrum suggests that IGE are not highly ionized in the optically thick ejecta. Therefore, the lack of any \ion{Co}{II} lines in the early NIR spectrum is likely the result of optical depth, rather than early abundance, which is suggested by our photometric measurements of optically thick ejecta. The appearance of the \ion{Co}{II} lines in the photosphere is somewhat at odds with the slow lightcurve evolution set by the long $\gamma$-ray escape time. However, if the ejecta is well-mixed as suggested by the full IR spectrum \cite{Kwok2025}, and as has been suggested for other SNe Iax \cite{Maeda2022}, then $^{56}$Ni products should be found in both the outer and inner ejecta, allowing the long $\gamma$-ray trapping.
We also note that \ion{Fe}{II} $\lambda 16907$ is slowly evolving in the NIR spectral series, possibly as a result of $^{56}$Co Decay, which would further suggest these Co features are the product of $^{56}$Ni decay.

By day +13.83 after \textit{B}-band maximum, the \ion{Ca}{II} NIR triplet becomes stronger in emission than absorption, and by day +25.51 we identify forbidden [\ion{Ca}{II}] $\lambda\lambda\,7291, 7323$. Such an early appearance of forbidden Ca has been seen in SN\,2008ha \citep{Valenti2009,Foley2009}, though the feature appears even earlier in \sn.

\subsubsection{Line Velocities}
In Extended Data Figure \ref{fig:velocities} we show the measured photospheric velocity $v_{\rm phot}$ of several prominent spectral lines. To measure this, we fit an appropriate phenomenological function to each line profile. The results of these fits are shown in Extended Data Figure \ref{fig:velocities_fit}. For \ion{C}{II} $\lambda 6578$, \ion{Si}{II} $\lambda6355$, and \ion{Fe}{II} $\lambda5169,\lambda6257$, we fit a simple negative Gaussian to the line absorption. The reason we chose a simple function is that these lines blend with other lines at later epochs, so complex functions failed to reproduce reliable measurements shortly after the B-band peak. Because of this blending, we only fit the absorption minimum velocity, since the edges are also not well-defined. We visually inspect each fit and remove results that are obviously contaminated by noise or line blending. For the \ion{Ca}{II} IR triplet, we fit a double P Cygni profile for the entire structure. This is done as follows: First, we remove a linear continuum. Then we fit 4 Gaussians: 2 are positive and centred at $\lambda8662$ and $\lambda\lambda8498,8542$, representing emission. The other two Gaussians are negative to the blue side of the emission, representing absorption. We use a single set of Gaussians for the blue lines because they are blended. We then measure the absorption minimum of $\lambda8662$, which is unblended. We find the fits to be more reliable than the absorption method. For the IR \ion{Co}{II} $15759,16064$ and $16361 \rm \AA$ lines, we fit an individual P Cygni profile to each line. To ensure consistency, we bind the three absorption minima velocity ($v_{\rm phot}$) and the FWHM velocity to be uniform across the three lines in each epoch. To check our fits, we also run these fits on the \jwst spectrum from ref. \cite{Kwok2025}, and find our results to be consistent with theirs. Our results show that the early spectra IME lines (\ion{Si}{II} and \ion{Ca}{II}) and IGE \ion{Fe}{II} exhibit a higher expansion velocity ($v_{\rm phot}\sim3500\,\rm km\,\rm s^{-1}$) than LME \ion{C}{II}  ($v_{\rm phot}\sim2800\,\rm km\,\rm s^{-1}$). This early difference could be because these are stronger transitions and hence sample the very outer ejecta early on. Later, however, \ion{C}{II}, \ion{Si}{II} and IGE lines show consistent velocities, declining to a low expansion velocity of $v_{\rm phot}\lesssim1000\,\rm km\,\rm s^{-1}$ by day 30. However, \ion{C}{II} declines in velocity much faster, and \ion{Ca}{II} remains at a higher velocity ($v\sim1500 \,\rm km \,s^{-1}$) for a longer time. 

The uniform velocity of LME, IME, and IGE, except Ca, is consistent with the IR spectra presented in \cite{Kwok2025} and suggests well-mixed ejecta. However, the Ca NIR triplet remains an outlier in that regard. One simple explanation for the high velocity of \ion{Ca}{II} is that it has a large optical depth due to its very strong transition. Therefore, it measures the velocity space more thoroughly, representing the fastest ejecta. This ejecta would also be thinner than the inner ejecta represented by the slower-expanding IGE absorption zones.

The appearance of forbidden \ion{Ca}{II} at an early epoch also appears to be at odds with the optical depth probed by the slow declining lightcurve and $t_{0}$ $\gamma$-ray escape time. However, both probes measure the effective optical depth in the Co-rich layers. Therefore, we find further evidence that some Ca lines form further away in a less dense part of the ejecta. If Ca lines only form at the outer ejecta, this could suggest it is richer in IME compared to the inner ejecta. 
An upper ejecta structure has been seen in other SNe Iax \cite{Maeda2022}. In some cases, the upper ejecta is optically thick enough to create two $^{56}$Ni decay components \cite{Kawabata2018,Kawabata2021}. In \sn we do not find any upper ejecta that affects the lightcurve, possibly due to a low amount of mass or the overall slow velocity regime of the entire ejecta.

\subsection{Powering mechanism}\label{sec:power}
SNe Ia are powered by the decay of radioactive nickel \citep{Arnett1979,Arnett1982}. While SNe Iax are thought to arise from the deflagration, rather than detonation of WD material \cite{Fink2014}, their lightcurves should still follow the same thermonuclear principles.
Therefore, to understand the energetics, it is imperative to probe the amount of $^{56}$Ni synthesised by the SN explosion. As the SN ejecta become optically thin, we expect the bolometric luminosity to equate the deposited energy by $\gamma$-rays and positrons emitted by the decay. These depositions are described by the following set of equations \citep{Swartz1995,Junde2011}:
\begin{equation}
    Q_{\gamma}(t) = \frac{M_{\rm Ni_{56}}}{{\rm M_{\odot}}}[6.54e^{-\frac{t}{8.76\,\rm d}}+1.38e^{-\frac{t}{111.4\,\rm d}}]\times10^{43}\,\rm erg\,\rm s^{-1}
\end{equation}
\begin{equation}
    Q_{pos}(t) = 4.64\frac{\rm M_{Ni_{56}}}{M_{\odot}}[e^{-\frac{t}{111.4\,\rm d}}-e^{-\frac{t}{8.76\,\rm d}}]\times10^{41}\,\rm erg\,\rm s^{-1}
\end{equation}
Where $Q_{\gamma}$ is the $\gamma$-ray deposition, $Q_{pos}$ is the positron deposition and $M_{\rm Ni_{56}}$ is the synthesised Nickel mass.

As the ejecta become optically thin, more $\gamma$-ray photons escape, thus reducing the efficiency of $\gamma$-ray deposition \citep{Jeffery1999}. The fraction of depositing $\gamma$-ray photons $f_{\rm dep}$ scales as $f_{\rm dep}=1 -e^{(\frac{t_{0}}{t})^{2}}$ where $t_{0}$ is the $\gamma$-ray escape time. Therefore, the total deposition is given by:
\begin{equation}
    Q_{\rm dep}=Q_{\gamma} f_{dep}+Q_{pos}
\end{equation} \label{eq:dep}

To measure the Nickel mass of \sn, we fit Eq. (3) to the late bolometric lightcurve ($t>45$ days after explosion) and up to the our last \textit{B}-band measurement ($57$ days after explosion), corresponding to a time where the SN should be optically thin and we have good photometric coverage. We find the best fit to the lightcurve to be $M_{\rm ^{56}Ni}=3.67_{-0.45}^{+0.1}\times10^{-3}\,\rm M_{\odot}$ in line with other faint SNe Iax \citep{Karambelkar2021}. We further find the best fit for $t_{0}$ to be $\approx70$ days. To confirm this, we probe the $t_{0}$ parameter using another method, using the entire lightcurve, including its peak.
The internal energy of the low mass ejecta, having gone through significant adiabatic losses (since the progenitor is compact), is negligible compared to the contribution of radioactive decay energy. Therefore, one can measure the $t_{0}$ parameter independently from the nickel mass, using conservation of energy and based on the Katz integral \citep{Katz2013,Kushnir2013}, whereby:
\begin{equation}
    \frac{L(t)}{\int_{0}^{t}L(t')t'dt} = \frac{Q(t)}{\int_{0}^{t}Q(t',t_{0})t'dt}
\end{equation}\label{eq:katz_integral}
This equation holds for the late lightcurve, where opacity effects are averaged out. This is seen in Figure \ref{fig:Thermo_figure} panel c, as the convergence of both the nominators and denominators in both sides of equation (4) reaching unity.

We solve Eq. (4) for 57 days, and find the best fit $t_{0}$ to be 83.05 days. We then probe the parameter space by running a basic 1-d Markov chain Monte Carlo (MCMC) algorithm using the inferred bolometric lightcurve error. We find a median value of $t_{0}=83.05^{+14.93}_{-15.71}$ days ($1 \sigma$ errors) days and that $t_{0} > 48.2$ days with a $3\sigma$ confidence. This confirms that \sn has an unusually high $\gamma$-ray escape time, suggesting the SN ejecta remains optically thick to $\gamma$-rays for a long duration after explosion. We further find that $^{56}\rm Ni$ accounts for 99.98\% of the SN luminosity for the best fit $t_{0}$ value, ruling out any additional energy source.

We find that the low ejecta expansion velocity (as measured by slow photospheric velocities) provides a good explanation to the slow lightcurve evolution. The $\gamma$-ray escape time, $t_{\gamma}$, beyond which $\gamma$-ray deposition becomes inefficient is determined by the plasma column density $\Sigma$ (i.e the ejecta density at the line-of-sight) and an effective $\gamma$-ray opacity $\kappa_{\gamma}\approx0.025$ that has very weak dependence on chemical composition \cite{Guttman2024}. This dependence is given by:
\begin{equation}
    t_{\gamma}=\sqrt{\kappa_{\gamma}\,\Sigma(t)\,t^{2}}
\end{equation}
Where $t$ is the time since explosion.
For a homologous expanding ejecta, $\Sigma$ is given by:
\begin{equation}
    \Sigma (t) = \frac{M_{ej}}{4\pi\,v_{ej}\,t^{2}}
\end{equation}
Where $M_{ej}$ is the ejecta mass and $v_{ej}$ is the ejecta velocity.
Therefore, the ejecta mass ejected by a thermonuclear explosion can be assessed as:
\begin{equation}
    M_{ej}=\frac{4\pi\,t_{\gamma}^{2}\,v_{ej}^2}{\kappa_{\gamma}}
\end{equation}

For $t_{\gamma}=80$ days as inferred from the Katz integral method and an ejecta velocity of $v\approx1,000\rm\,km\,s^{-1}$ we find that an ejecta mass of $M_{ej}\approx0.12\rm\,M_{\odot}$ is produced.

To check that the ejecta mass is consistent with other measurements, we assess it using its bolometric rise time. This provides an independent measurement of the ejecta mass. The ejecta mass of a thermonuclear explosion can be assessed using a scaling relation\cite{Foley2009,Pinto2000}:
\begin{equation}
    M_{ej}=0.16\,(\frac{t_{r}}{10 \rm\,days})(\frac{0.1\rm\,cm^{2}\,g^{-1}}{\kappa_{\rm opt}})(\frac{v_{ej}}{2\times10^{8}\rm\,cm})\,M_{\odot}
\end{equation}
where $t_{r}$ is the bolometric rise time and $\kappa_{\rm opt}$ is the optical opacity. For \sn we measure a rise time of $t_{r}\approx11.3$ days. Assuming an opacity of $0.1\rm\,cm^{2}\,g^{-1}$, we find the ejecta mss to be $M_{ej}\approx0.1\rm\,M_{\odot}$.
While detailed modelling is needed to precisely calculate the ejecta mass, the agreement between these two assessments strengthens the hypothesis that the slow-expanding ejecta is the source of long $\gamma$-ray escape time duration, and hence its slow bolometric decline.

\subsubsection{Bound remnant and late-time lightcurve}
Type SNe Iax often display a late-time luminosity excess exceeding that of pure thermonuclear explosion models \cite{Kawabata2018,Kawabata2021}. This excess is often found to begin $>50$ days past their explosion, and can potentially last for years \cite{Foley2014,McCully2022,Schwab2025}. A gravitationally bound remnant, left behind the weak explosion, is often credited to create this discrepancy, possibly due to $^{56}$Ni deep within the bound remnant or secondary ejecta produced by it \cite{Maeda2022}. Due to the faintness and since \sn's field was setting, our multiband lightcurve coverage ended too early to find any discrepancy from a single thermonuclear explosion.

However, as the field with \sn became observable again, we obtained further observations of \sn. Interestingly, \sn was not detected in bands other than $i$-band in regular LCO images, while deep images taken with EFOSC2 and Magellen/IMACS provided detections in the $z$, $r$ and $g$ -bands as well. In all cases, the $i$-band was significantly brighter than the other bands. This is likely caused by strong lines in the SN spectra, which can be attributed to the forbidden \ion{Ca}{II} NIR doublet. Interestingly, at no late epoch did the optical bands (redder than $i$-band) cover more than $10\%$ of the total luminosity predicted by Nickel decay, when integrated together and compared to the expected luminosity from $^{56}$Ni decay. This means that at late times, most of the luminosity released by \sn is emitted in the IR. To confirm this, we integrate the flux from a \jwst spectrum obtained at $+194$ days post-peak, covering $0.7$--$13\,\rm\mu m$, obtained through a director discretionary program (PI: Baron). We find that more than $50\%$ of the flux is emitted at wavelengths longer than $2\rm \mu m$ at this epoch. This suggests the \sn caused dust formation at these later epochs, shifting the SED to the infrared.

Using late-time data, we find no additional emission from the bound remnant. The flux from the \jwst spectrum matches the expected flux from the late-time $^{56}$Ni decay tail. Furthermore, by correcting the late $i$-band assuming they account $6 \%$ of the \sn flux, we can reconstruct a lightcurve that follows the $^{56}$Ni deposition tail. Since the $i$-band decline rate follows that of a single $^{56}$Ni tail, it confirms that no excess energy source exists at these epochs for \sn. It should be noted, however, that at later epochs this correction fails, as expected if the temperature further drops or the effects of dust formation accumulate, decreasing the $i$-band fraction of the total flux. We therefore cannot construct the decline rate longer than $209$ days, such that excess flux from the remnant would be possible at these later epochs. Regardless, such flux is only expected in the deep IR, necessitating further \jwst observations to detect.

\clearpage

\subsection{Extended Data Figure}
\setcounter{figure}{0}

\begin{figure}[h!]
    \captionsetup{name=\extmattwo Figure}
    \centering
    \includegraphics[width=\linewidth]{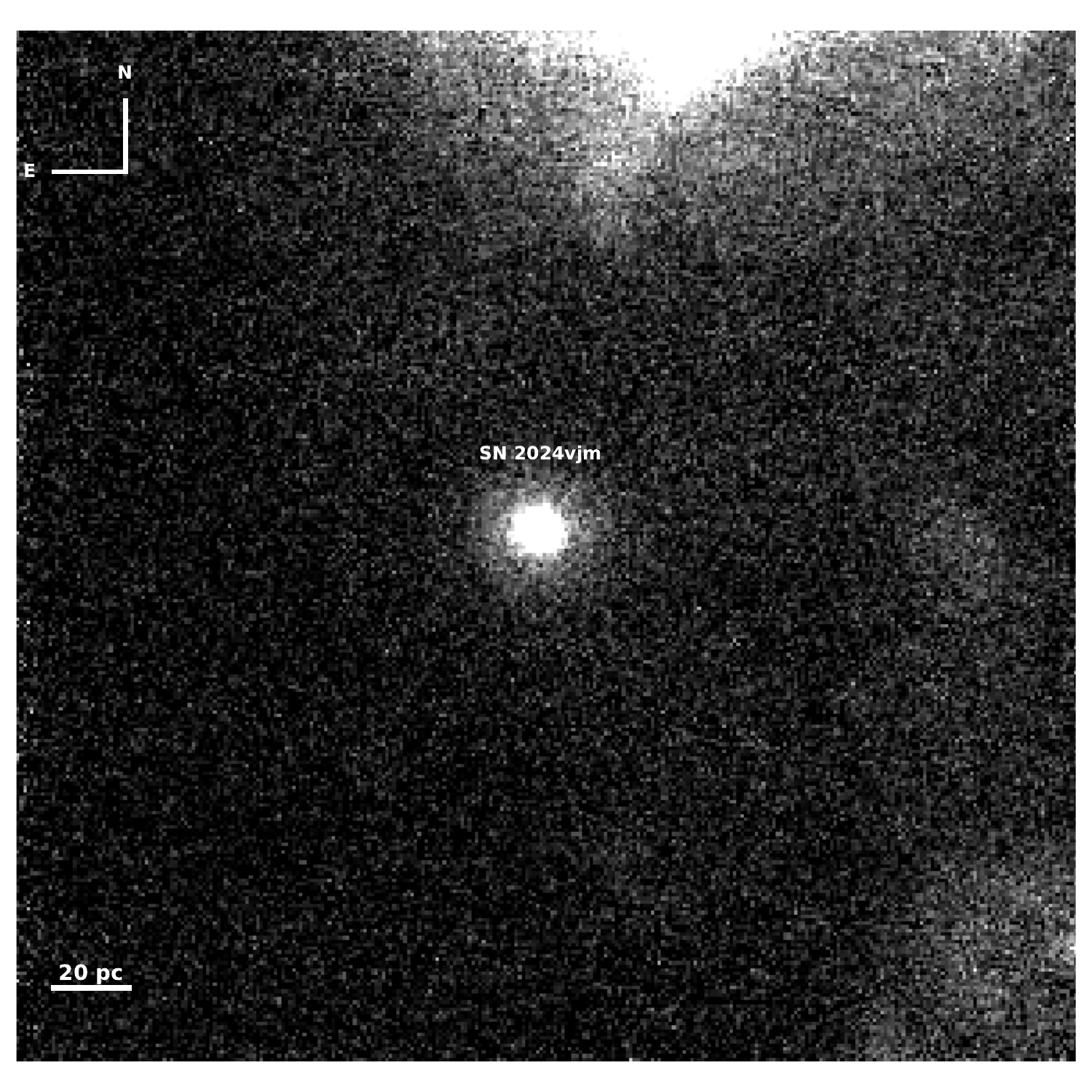}
    \caption{\sn is surrounded by star-formation. The host redshifted $\rm H\alpha$ slice of the MUSE datacube is shown. \sn is clearly visible at the centre and is marked by its name. The edges of two large star-forming regions are seen to the north ($\sim 80$pc) and south-west ($\sim140$ pc) of the SN. However, the SN lies away from large star-forming regions. The emission lines produced by the northern star-forming region are consistent with a solar abundence of metals.}
    \label{fig:MUSE_Ha}
\end{figure}

\begin{figure}[h!]
    \centering
    \captionsetup{name=\extmattwo Figure}
    \includegraphics[width=\linewidth]{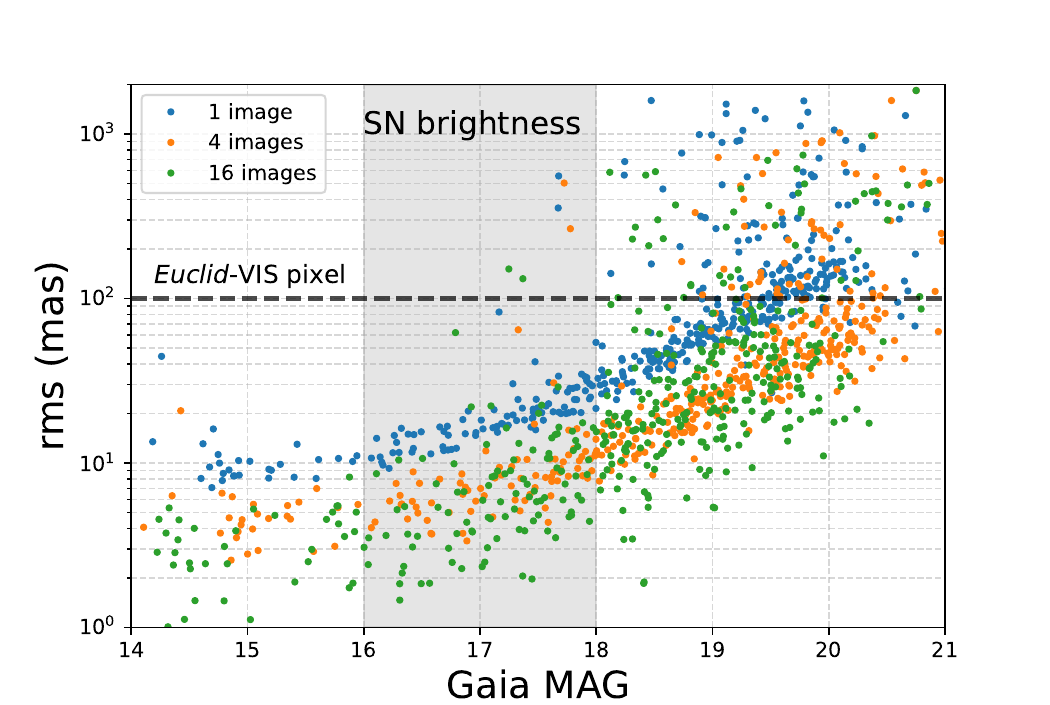}
    \caption{Binning seeing-limited BG images improves the astrometric accuracy. Median astrometric position of identified Gaia sources within BG $i$-band images in single images (blue), four binned images (orange) and sixteen binned images (green). By combining more images the astrometric noise decreases by roughly the square root of the number of images. This demonstrates that the mean astrometric points are largely independent, and therefore, averaging multiple data points can achieve better accuracy. A single \euclid-VIS pixel is marked at $0.1''$ with a striped line. The brightness of \sn within the $i$-band images is marked with a shaded grey region. }
    \label{fig:astrometry_i_band}
\end{figure}

\begin{figure}[h!]
    \centering
    \captionsetup{name=\extmattwo Figure}
    \includegraphics[width=\linewidth]{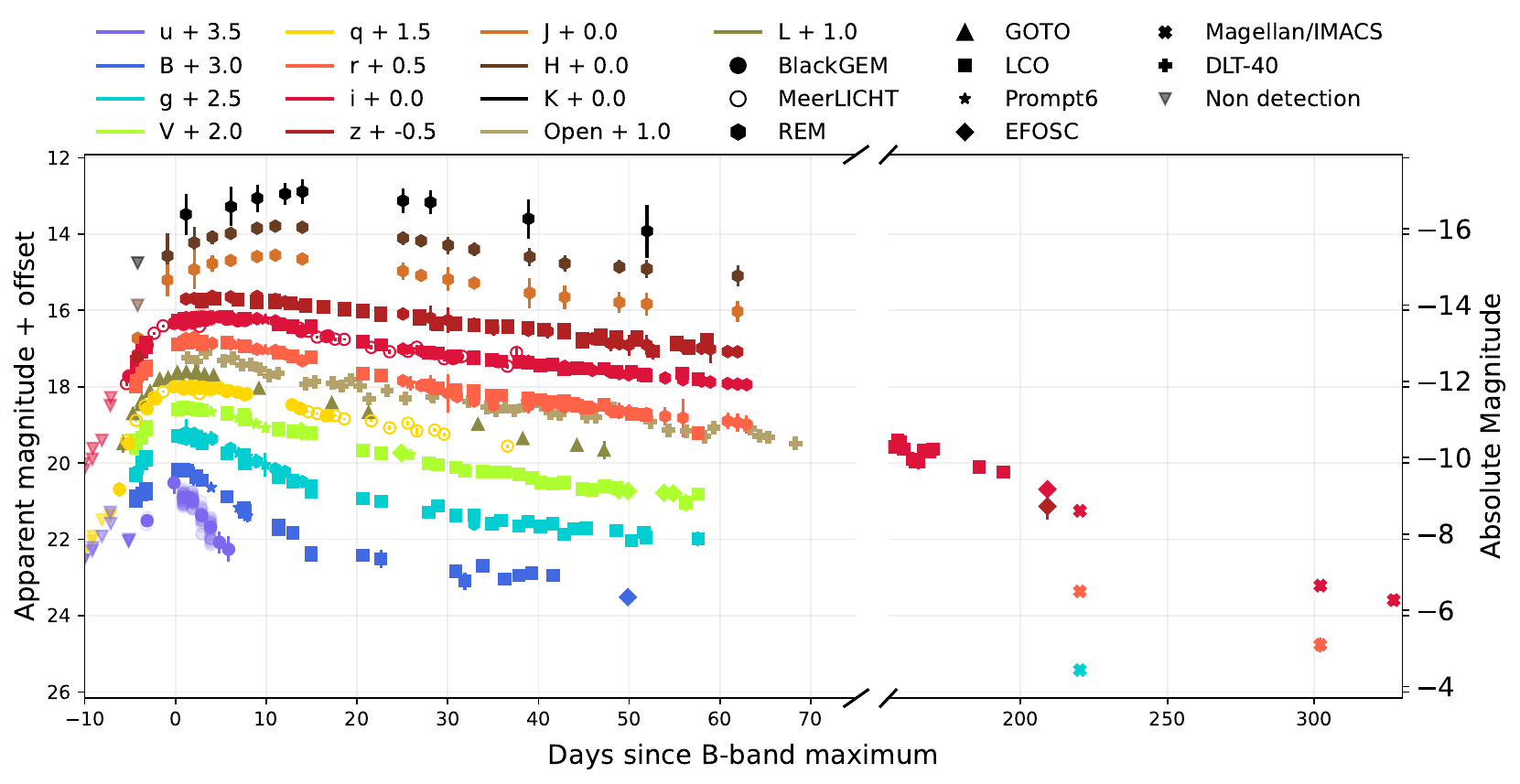}
    \caption{Mult$i$-band lightcurve of \sn. Different filter bands are plotted with vertical offsets for clarity. $BVJHK$ bands are shown in the Vega magnitude system following convention, while all other bands are in the AB magnitude system. We show non-detections up to the first detection in each band in triangles. Individual unbinned measurements are shown with semi-transparent markers, while binned data points are shown in opaque markers. All other data are displayed with marker styles and colours corresponding to their respective telescope and band. The gap in observations is due to the field setting behind the sun. Error bars represent $1\sigma$ uncertainties.}
    \label{fig:lightcurve}
\end{figure}

\begin{table}[ht]
    \centering
    \captionsetup{name=\extmattwo Table}
    \caption{Lightcurve parameters of \sn. All values are extinction extinction corrected. \sn is possibly the faintest SN observed to date. The errors are of $1\,\sigma$.}
    \label{tab:lc_prop}
    \footnotesize 
    \begin{tabular}{l c c c c}
        \hline\hline
        Filter & Maximum light (MJD) & Peak apparent magnitude & Peak absolute magnitude & $\Delta M_{15}$ \\
        \hline
    u (BG) & 60572.64$\pm$0.41 & $17.39\pm0.10$ & $-12.47 \pm 0.14$ & $1.15\pm0.56$\\
    B & 60572.29$\pm$0.17 & $17.18\pm0.03$ & $-12.68 \pm 0.10$ & $1.85\pm0.05$\\
    g & 60572.50$\pm$0.03 & $16.80\pm0.02$ & $-13.06 \pm 0.10$ & $1.32\pm0.04$\\
    V & 60573.23$\pm$0.14 & $16.56\pm0.01$ & $-13.30 \pm 0.10$ & $0.74\pm0.06$\\
    q & 60573.74$\pm$0.04 & $16.53\pm0.04$ & $-13.33 \pm 0.11$ & $0.71\pm0.04$\\
    r & 60574.43$\pm$0.07 & $16.33\pm0.01$ & $-13.53 \pm 0.10$ & $0.58\pm0.13$\\
    i & 60577.63$\pm$0.21 & $16.17\pm0.01$ & $-13.69 \pm 0.10$ & $0.57\pm0.02$\\
    z & 60576.86$\pm$0.34 & $16.20\pm0.01$ & $-13.66 \pm 0.10$ & $0.31\pm0.03$\\
    J & 60582.35$\pm$0.83 & $14.61\pm0.10$ & $-15.25 \pm 0.14$ & $0.38\pm0.14$\\
    H & 60583.40$\pm$0.22 & $13.83\pm0.07$ & $-16.03 \pm 0.12$ & $0.32\pm0.11$\\
    K & 60586.31$\pm$0.15 & $13.04\pm0.19$ & $-16.82 \pm 0.22$ & $0.20\pm0.29$\\
        \hline
    \end{tabular}
\end{table}

\begin{figure}[h!]
    \centering
    \captionsetup{name=\extmattwo Figure}
    \includegraphics[width=\linewidth]{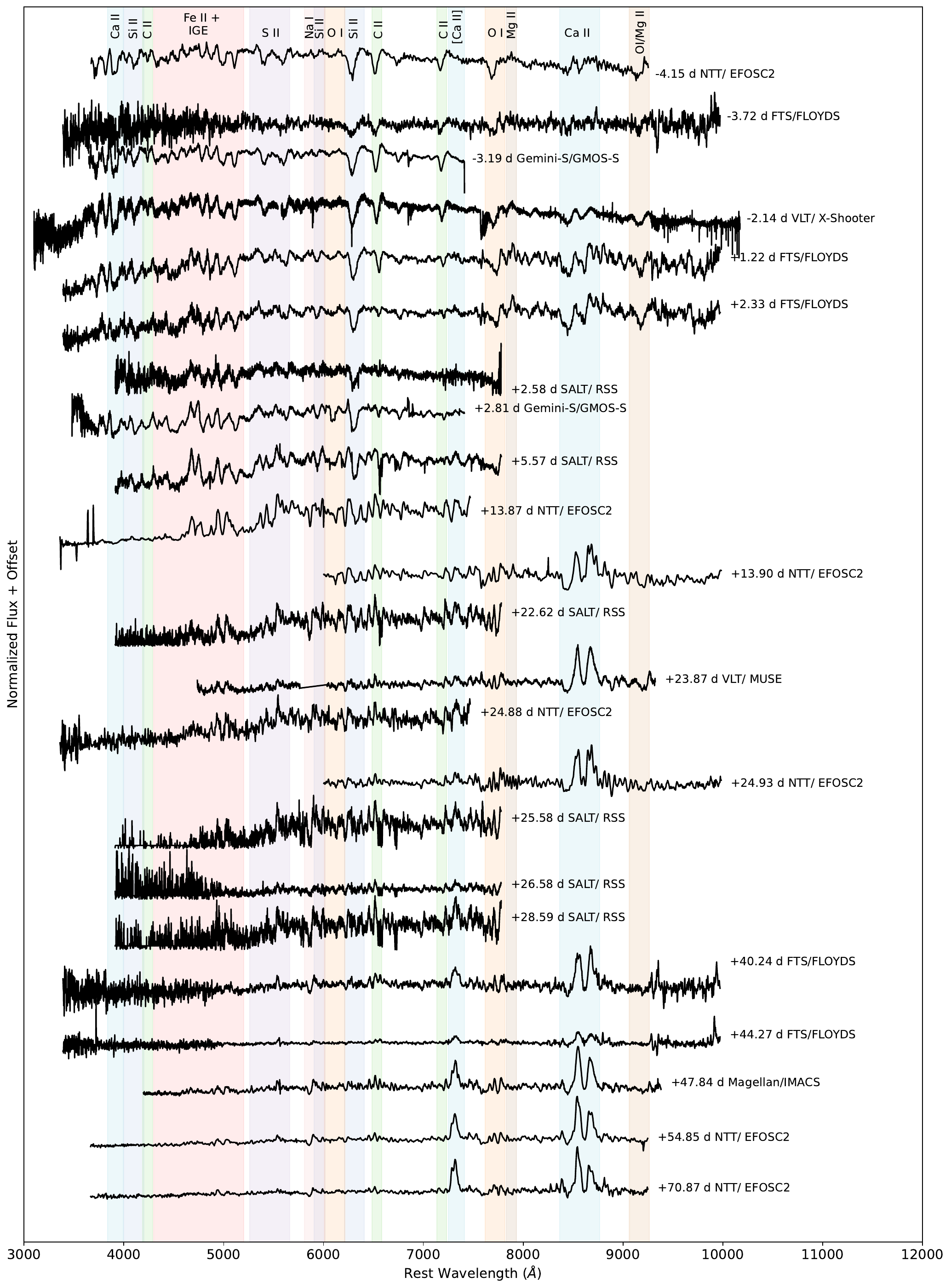}
    \caption{Optical spectra of SN\,2024vjm. The spectra are ordered in phase from \textit{B}-band maximum and are corrected for extinction. Prominent features are marked in shaded colour.}
    \label{fig:spec}
\end{figure}

\begin{figure}
    \centering
    \captionsetup{name=\extmattwo Figure}
    \includegraphics[width=\linewidth]{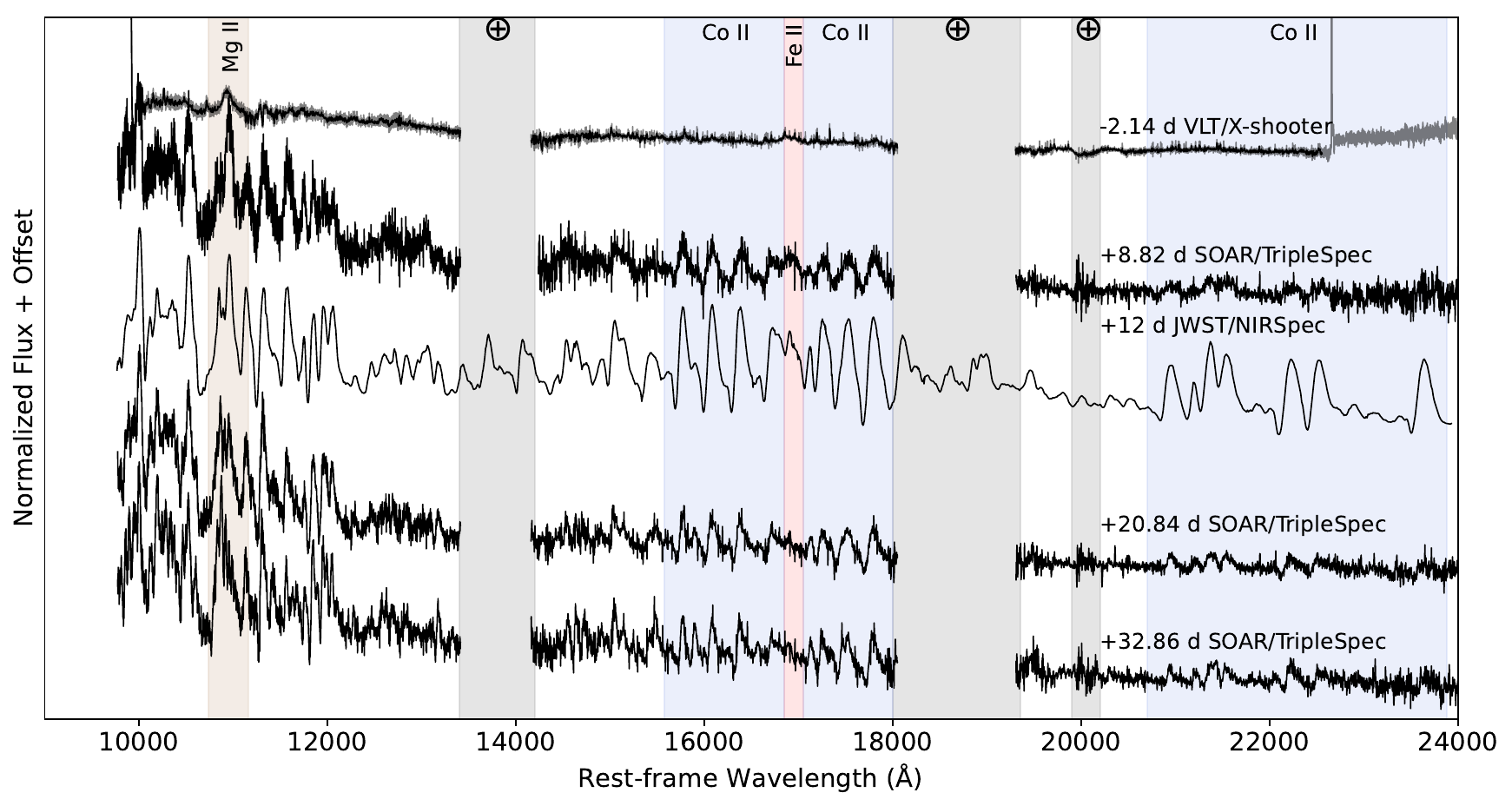}
    \caption{NIR Spectra of \sn. The NIR section of the +12 d \jwst spectrum from \cite{Kwok2025} is shown along with spectra obtained for this study. While early on the spectrum is almost featureless, barring \ion{Mg}{II} and a possible single \ion{Fe}{II} line, later epochs show striking \ion{Co}{II} P Cygni profiles throughout the H and K band.}
    \label{fig:NIR_spec}
\end{figure}

\begin{figure}
    \centering
    \captionsetup{name=\extmattwo Figure}
    \includegraphics[width=\linewidth]{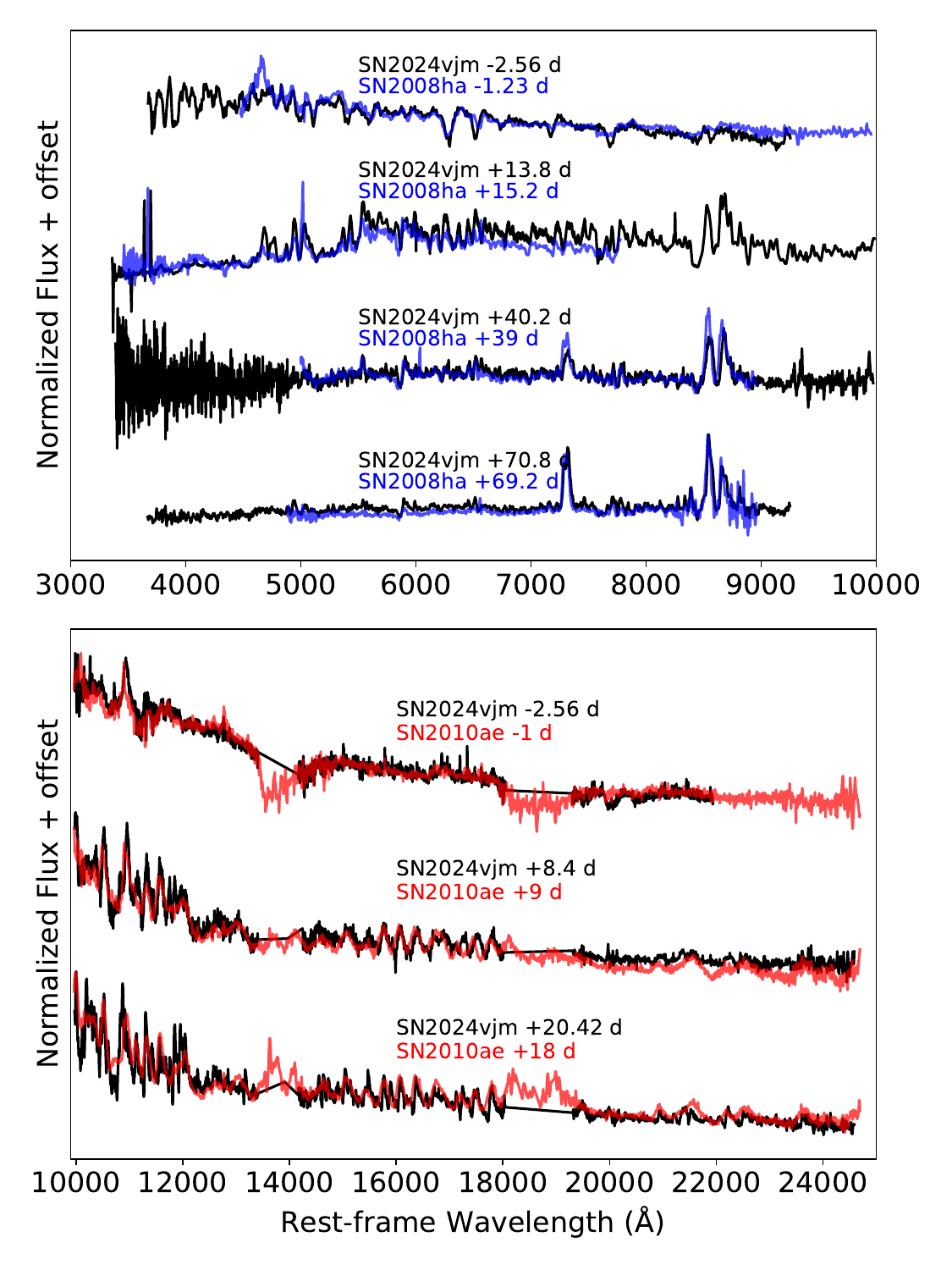}
    \caption{\sn shows spectral similarity to other faint SNe Iax. We compare the normalised spectra of \sn to faint SNe Iax SN\,2008ha (blue, optical) and SN\,2010ae (red, NIR). The spectral evolution of \sn is similar to the other faint SNe Iax per epoch, although its expansion velocity is smaller than the other two faint SNe.}
    \label{fig:spec_comparison}
\end{figure}

\begin{figure}
    \centering
    \captionsetup{name=\extmattwo Figure}
    \includegraphics[width=\linewidth]{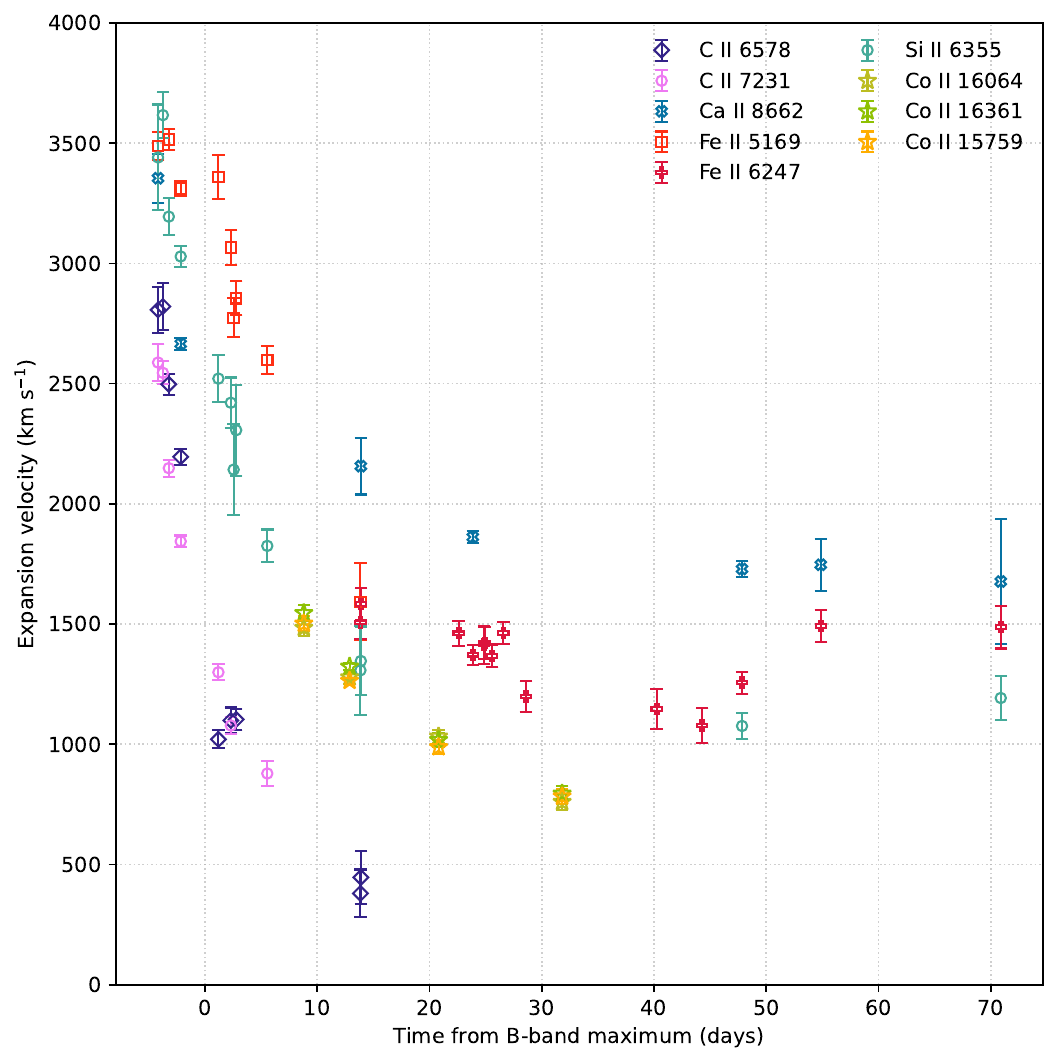}
    \caption{The ejecta of \sn was slowly expanding. Absorption line velocities from SN\,2024vjm's spectra are shown in different phases. Different lines are marked with distinct colours. We initially measure similar velocities in all lines; however, we find that Ca diverges from the IGE at later epochs, maintaining a larger expansion velocity. Error bars represent $1\sigma$ uncertainties.}
    \label{fig:velocities}
\end{figure}

\begin{figure}
    \centering
    \captionsetup{name=\extmattwo Figure}
    \includegraphics[width=\linewidth]{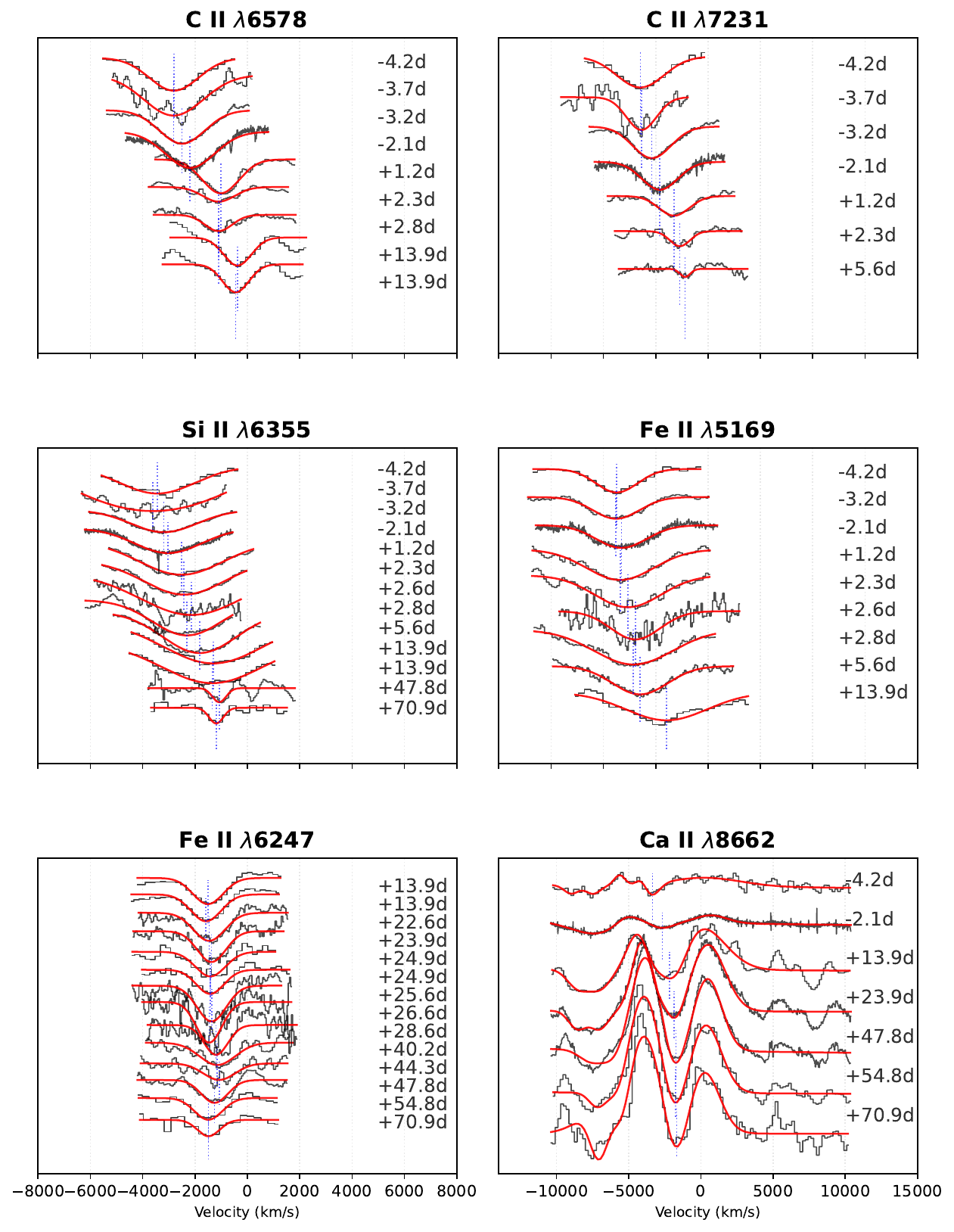}
    \caption{Velocity fits to prominent spectral lines of \sn. The fit (red) to the original spectral data (black) shows that the fits match the original shape well. The absorption minima of the different lines are marked (blue-striped line), showing that all spectral features decline in early epochs.}
    \label{fig:velocities_fit}
\end{figure}

\clearpage
\section{Declarations}\label{decl}

\backmatter

\bmhead{Supplementary information}
Supplementary Information is available for this paper.

\bmhead{Correspondence}
Correspondence and requests for materials should be addressed to Erez~A.~Zimmerman~(email: erezimm@gmail.com).

\bmhead{Acknowledgements}
We thank Dr. Amir Sharon, Prof. Boaz Katz, Prof. Doron Kushnir and Prof. Eli Waxmann for meaningful discussions.
Based on observations with the BlackGEM telescope array. BlackGEM has been made possible with financial aid from Radboud University, the Netherlands Research School for Astronomy (NOVA), the KU Leuven Department of Physics and Astronomy, the Netherlands Organization for Scientific Research (NWO), the Research Foundation Flanders (FWO) under grant agreement G0A2917N (BlackGEM, with PIs C. Aerts \& G. Raskin and personnel C. Johnston and P. Ranaivomanana), the European Research Council. BlackGEM is a collaboration between NOVA, Radboud University, KU Leuven and the Universities of Manchester, Warwick, Hamburg, Potsdam, Barcelona, Durham, Tel Aviv and Valparaiso, and the Armagh Observatory and Planetarium, the University of California at Davis, the Danish Technical University, Texas Tech University, the Weizmann Institute, Las Cumbres Observatory, and the Hebrew University of Jerusalem. BlackGEM is hosted at the La Silla Observatory of the European Southern Observatory. 
Based on observations with the MeerLICHT telescope. The MeerLICHT telescope is operated at the SAAO Sutherland site and supported by a consortium consisting of Radboud University, the University of Cape Town, the University of Oxford, the University of Manchester, the University of Amsterdam and the South African Astronomical Observatory, on behalf of the National Science Foundation of South Africa. 
Based on observations collected at the European Southern Observatory under ESO programmes 1112.D-0334, 113.26B1, 114.28GZ and 1103.D-0328.
This work makes use of observations from the Las Cumbres Observatory network.
The LCO team is supported by NSF grants AST-2308113 and AST-1911151.
This research is based in part on data collected at the international Gemini Observatory within the framework of Subaru-Gemini time exchange program (S24B-041, GS-2024B-Q-101, GS-2024B-Q-102: PI, K. Maeda). We are honored and grateful for the opportunity of observing the Universe from Maunakea, which has the cultural, historical and natural significance in Hawaii.
This paper includes data gathered with the 6.5 meter Magellan Telescopes located at Las Campanas Observatory, Chile.
REM data were obtained under the program: REM AOT47-37 (ID 49337), "Characterising the elusive interacting transients", (PI: G. Valerin)
The Prompt telescopes belong to 'The Skynet Robotic Telescope Network which operates out of the University of North Carolina at Chapel Hill.
Supported by the National Science Foundation, North Carolina Space Grant, and the Mount Cuba Astronomical Foundation.'
This work has made use of the Early Release Observations (ERO) data from the {\it Euclid} mission of the European Space Agency (ESA), 2024,
\url{https://doi.org/10.57780/esa-qmocze3}. Euclid is a fully European mission, built and operated by ESA, with contributions from NASA. The Euclid Consortium is responsible for providing the scientific instruments and scientific data analysis. ESA selected Thales Alenia Space as prime contractor for the construction of the satellite and its Service Module, with Airbus Defence and Space chosen to develop the Payload Module, including the telescope. NASA provided the near-infrared detectors of the NISP instrument. Euclid is a medium-class mission in ESA’s Cosmic Vision Programme.
This work is based in part on observations made with the NASA/ESA/CSA James Webb Space Telescope. The data were obtained from the Mikulski Archive for Space Telescopes at the Space Telescope Science Institute, which is operated by the Association of Universities for Research in Astronomy, Inc., under NASA contract NAS 5-03127 for JWST. These observations are associated with program \#DD9231.
The authors acknowledge the team led by coPIs Eddie Baron and James M. DerKacy for developing their observing program with a zero-exclusive-access period.
This research made use of Photutils, an Astropy package for
detection and photometry of astronomical sources (Bradley et al. 2025)
PJG is supported by NRF SARChI grant 111692.
S.V. and the UC Davis time-domain research team acknowledge support by NSF grants AST-2407565.
PC acknowledges the support from the Zhejiang provincial top-level research support program. 
Time-domain research by the University of Arizona team and D.J.S. is supported by National Science Foundation (NSF) grants 2108032, 2308181, 2407566, and 2432036 and the Heising-Simons Foundation under grant \#2020-1864.
GL and SdW were supported by a research grant (VIL60862) from VILLUM FONDEN
This research was supported by Deutsche Forschungsgemeinschaft  (DFG, German Research Foundation) under Germany’s Excellence Strategy - EXC 2121 "Quantum Universe" – 390833306. Co-funded by the European Union (ERC, CompactBINARIES, 101078773). Views and opinions expressed are however those of the author(s) only and do not necessarily reflect those of the European Union or the European Research Council. Neither the European Union nor the granting authority can be held responsible for them.
N.B. acknowledges financial support from grant CEX2024-001451-M, funded by MICIU/AEI/10.13039/501100011033. N.B. acknowledges funding from the European Union (ERC, CET-3PO, 101042610). Views and opinions expressed are however those of the author(s) only and do not necessarily reflect those of the European Union or the European Research Council Executive Agency. Neither the European Union nor the granting authority can be held responsible for them.
LAK is supported by NASA through a Hubble Fellowship grant No. HF2-51579.001-A awarded by the Space Telescope Science Institute (STScI), which is operated by the Association of Universities for Research in Astronomy, Inc., for NASA, under contract NAS5-26555.
KM acknowledges funding from Horizon Europe ERC grant no. 101125877
M.A. is supported by NSF grant AST-2308113
K.M. acknowledges support from JSPS KAKENHI grant (JP24KK0070, JP24H01810) and JSPS Bilateral Joint Research Projects (JPJSBP120229923). 
H.K. was funded by the Research Council of Finland projects 324504, 328898, and 353019.
SM is funded by Leverhulme Trust grant RPG-2023-240.
AR acknowledges financial support from the GRAWITA Large Program Grant (PI P. D’Avanzo) and from the PRIN-INAF 2022 "Shedding light on the nature of gap transients: from the observations to the models".
AM gratefully acknowledges support from an STFC PhD studentship and the Faculty of Science and Technology at Lancaster University.
T.-W.C. acknowledges the financial support from the Yushan Fellow Program by the Ministry of Education, Taiwan (MOE-111-YSFMS-0008-001-P1) and the National Science and Technology Council, Taiwan (NSTC grant 114-2112-M-008-021-MY3).
W.Z. acknowledges support by the National Natural Science Foundation of China (Grant No. 12133005). 

\section*{Declarations}

\bmhead{Author Contribution}
\begin{itemize}

\item \textbf{Paper Writing} --- 
E.~A.~Zimmerman, A.~Gal-Yam

\item \textbf{Discussion and Interpretation} --- 
All authors contributed to discussions and interpretation.

\item \textbf{Discovery of \sn} ---
P.~J.~Groot

\item \textbf{Data analysis} ---
E.~A.~Zimmerman, E.~O.~Ofek

\item \textbf{Observations and Data Reduction} --- 
E.~A.~Zimmerman, A.~Pastorello, S.~Valenti, A.~P.~Ravi, P.~Chen, N.~Blagorodnova, M.~Wavasseur, M.~A.~Gómez-Muñoz, P.~M.~Vreeswijk, S.~de Wet, L.~.A.~Kwok, M.~Schwab, S.~W.~Jha, D.~Hiramatsu, J.~Li, H.~Kuncarayakti, K.~Maeda, G.~Pignata, A.~Reguitti, G.~Valerin, W.~Zang

\item\textbf{BlackGEM Collaboration} ---
P.~J.~Groot, J.~van Roestel, E.~A.~Zimmerman, A.~Gal-Yam, S.~Valenti, A.~P.~Ravi, N.~Blagorodnova, M.~Wavasseur, M.~A.~Gómez-Muñoz, H.~Tranin, P.~M.~Vreeswijk, S.~de Wet, G.~Leloudas, E.~Stringer, T.~Kupfer, S.~Bloemen, D.~L.~A.Pieterse, F.~Stoppa

\item\textbf{ePESSTO+ Collaboration} ---
A.~Pastorello, J.~P.~Anderson, T.~-W.~Chen, K.~Maguire, M.~Della Valle, G.~Dimitriadis, ,M.~Gromadzki, J.~H.~Gillanders, J.~D.~Lyman, M.~R.~Magee, A.~Milligan G.~Pignata, A.~Reguitti, R.~P.~Santos, S.~Srivastav, G.~Valerin

\item\textbf{LCO Collaboration} ---
D.~A.~Howell, S.~W.~Jha, M.~Schwab, M.~Andrews, D.~Hiramatsu, S.~Moran, Y.~Ni, X.~Wang, J.~R.~Farah, K.~Wynn

\item\textbf{DLT-40 Collaboration} ---
D.~J.~Sand, S.~Valenti, A.~P.~Ravi, J.~Pearson, Y.~Dong, M.~Shrestha, K.~A.~Bostroem, N.~M.~Retamal, D.~Janzen, D.~E.~Reichart, B.~Subrayan, E.~Hoang, J.~Andrews, D.~Mehta

\end{itemize}

\bmhead{Competing Interests}
The authors declare that they have no competing financial interests.
\bmhead{Data Availability Statement}
Photometry and spectra used in this study will be made available on WISeREP\cite{yaron2012}. A log of the available spectra can be found in Supplementary Table \ref{tab:spectra}. 

\bmhead{Code Availability Statement}
All scripts used to conduct the analyses presented in this paper are available from the corresponding author upon request.
Relevant software sources have been provided in the text, web locations provided as references, and are publicly available.

\clearpage

\section{Supplementary Methods}\label{supp}

\subsection{Photometric reductions}\label{sec:phot_red}
Data was obtained with the facilities listed below:
\begin{itemize}
    \item BlackGEM and MeerLICHT -- BG photometry in \textit{uqi} bands was obtained by the BG LTS and fast-synoptic survey, with an intra-night cadence. Additional \textit{uqi} images were obtained by MeerLICHT (ML), which is the BG sister project at the South African Astronomical Observatory (SAAO). Observations were performed until the SN field began setting. The photometry was reduced by the BlackGEM \texttt{BlackBOX} pipeline \citep{Vreeswijk2026}, which performs the astrometric and photometric calibration based on Gaia DR3 data \cite{GAIA+2016_GAIA_mission,GAIA+2022yCat_GAIA_DR3_MainSourcesCatalog}, determines the PSF as a function of pixel position, and performs image subtraction with respect to a pre-built reference image closely following \texttt{ZOGY} \citep{ZOGY}, which includes an estimate of the transient PSF photometry. Using the so-called Scorr (significance) image, objects with $S/N \geq 6$ are selected as transient candidates. To identify detections with a lower $S/N>3$ directly at the SN location, we ran a forced-photometry pipeline on the BG images. This resulted in a single ($5\,\sigma$) pre-discovery $q$-band detection at an absolute magnitude of $M_{q} = -9.56\pm0.61$ mag (apparent $m_{q}=19.92\pm0.19$ mag) a day before detection ($MJD=60566.00$).
    
    \item LCO -- \textit{BVgriz} photometry was obtained by the Global Supernova Project (GSP) through the Las Cumbres Observatory \citep[LCO;][]{Brown2013} 1 meter telescope network. We performed PSF photometry with the {\tt daophot} task in IRAF on the reduced images with the LCOGT/BANZAI pipeline \citep{McCully2018_lcogt_Banzai} without image subtraction. We utilize the ATLAS All-sky Stellar Reference Catalog (ATLAS-REFCAT2; \citealt{Tonry2018_refcat2}) to derive the photometric zero-point. Before being used for photometric calibrations of our target, the ATLAS-REFCAT2 magnitudes of the reference stars in the fields are first converted into Johnson $BV$ and Sloan-$griz$ bands adopting the following transformations given in \cite {Tonry2012}. We also performed image subtraction on some selected images in the $r$ band using the image taken on 5 June 2025 as a template image, then obtained PSF photometry on the subtracted images. We obtained photometry results consistent with those without image subtraction, confirming that image subtraction is not necessary for LCOGT images due to a relatively faint and smooth host galaxy background. All these photometry procedures were performed using the {\tt pmpyeasy} pipeline \citep{Chen2022_pmpyeasy}.
    
    \item REM -- Additional photometry was obtained by the 60-cm Rapid Eye Mount (REM) telescope hosted in the La-Silla observatory. Thanks to a dichroic, simultaneous optical (\textit{griz}) and near-infrared (JHK) observations can be obtained using two cameras, ROSS2 for the optical and REMIR for the NIR. REM/ROSS2 images were treated with standard reduction steps, such as bias and flat field corrections, and final trimming. Individual images were then combined in a single (stacked) image to increase the signal-to-noise ratio. For the REM/REMIR images, pre-reduction only consisted in the subtraction of a sky frame before combining the images to obtain a final stack. Photometric measurements were performed using the  \texttt{ECsnoopy} pipeline\footnote{\texttt{ECsnoopy} is a package for supernova photometry using PSF fitting and/or template subtraction developed by E. Cappellaro. A package description can be found at the website: http: //sngroup.oapd.inaf.it/ecECsnoopy.html}.
    \texttt{ECsnoopy} allows the user to perform an accurate astrometric calibration of the stacked images and to measure the instrumental PSF-fitting photometry of the target. For the optical images, the magnitude measurements were performed after the subtraction of a host galaxy template\footnote{Skymapper \textit{griz} templates were adopted.}.
    No template subtraction was applied to the near-infrared images, given the modest host galaxy contamination in the SN region. The final photometric calibration was performed accounting for the zero point and colour-term corrections for each instrumental configuration, and making use of secondary standards from the Skymapper (\textit{griz}) and the 2MASS (JHK) catalogues.
    
    The first epoch of data was obtained through OPTICON program 24B042 (PI: Blagorodnova), while the rest of the data were obtained through REM program AOT47-37 (ID 49337; PI: G. Valerin).
    
    \item Prompt-6 -- Further \textit{BVgriz} photometry was obtained by the 0.41m Prompt-6 telescope hosted at the Cerro-Tololo Inter-American Observatory (CTIO; Chile) equipped with an FLI CCD. The data were taken under the CNTAC (Chilean National Telescope Allocation Committee) time and were pre-reduced for bias and flats.
    The images were reduced using the \texttt{ECsnoopy} software in a similar fashion to that of the REM photometry. To construct a magnitude catalogue with secondary standards in the Johnson-Bessell system, we transformed Skymapper \textit{griz} magnitudes into Johnson-Bessell ones using the transformation relations of \cite{Chonis2008}.
    \item DLT 40  -- Follow up photometry was taken by the DLT 40  \citep{Tartaglia2018} survey using the PROMPT 0.4 m telescope at the Cerro Tololo Inter-American Observatory (CTIO), Chile. The observations were taken with no filter, and were calibrated to the APASS r-band catalogue \citep{Tartaglia2018}. The observations were grouped into one-day bins to improve the signal.
    \item GOTO -- Data from The Gravitational-wave Optical Transient Observer (GOTO) \cite{Steeghs2022} were obtained in the primary GOTO-L filter ($\sim400-700$\,nm). Following discovery announcement by the BlackGEM collaboration, a recent detection in GOTO's regular all-sky survey was found, and nightly-cadence observations scheduled to capture the rise and peak of SN\,2024vjm. Images had CCD reduction, calibration and photometry performed in real-time using the GOTO transient pipeline detailed in ref. \cite{Lyman2026}. Astrometric and photometric calibration were performed using Gaia DR3 data \cite{GAIA+2016_GAIA_mission,GAIA+2022yCat_GAIA_DR3_MainSourcesCatalog} and ATLAS REFCAT2 \cite{Tonry2018}. Difference image analysis was performed by a multi-threaded version of HOTPANTS \cite{Becker2015}, using historical templates taken by GOTO of the same region of sky. Forced aperture photometry at the position of SN\,2024vjm was performed to recover the final GOTO light curve.
    
    \item NTT/EFOSC2 -- The aquisition images taken before obtaining NTT/EFOSC2 spectra were reduced using the \texttt{ECsnoopy} pipeline in a similar fashion to that of the REM images described above. Additional late-time photometry was reduced using \textit{AstroPack/MAATv2} using the DELVE DR2 star-catalog \cite{Drlica2022} as reference.
\end{itemize}

\subsection{Lightcurve interpolation} \label{app:lc}
As described in \extmat \ref{sec:photometry}, to compute the lightcurve properties, we interpolated the different filter-band curves using a custom GP fitter based on the \texttt{sklearn} \citep{scikit-learn} \texttt{Python} Package. We run this interpolation in flux-space.
To instigate a rising interpolation function in the early data, we added $0$ flux ``ghost points" at the inferred explosion time. This anchors the GP to zero flux before the explosion and prevents unphysical extrapolation to pre-explosion time.
Because the temporal behavior of the light curve differs between the rapid rise to maximum and the smoother post-maximum decline, we employ two distinct GP kernels. For the early data (up to $\sim$10 days after explosion), we use \texttt{Matern} kernel (with a parameter of $\nu=2.5$), and a short characteristic length scale (of $2$--$12$ days), allowing the GP flexibility during the rise time. For later epochs, we use a long-scale squared-exponential (\texttt{RBF}) component (with a fixed $10$ day kernel) plus a linear \texttt{DotProduct} component. To ensure continuity, the early- and late-phase GPs are blended smoothly using a logistic transition function centred at 10\,d after explosion. The final interpolant is the weighted combination of the two GP predictions, with uncertainties propagated in quadrature. 
To mitigate GP overconfidence in sparsely sampled parts of the data, we further inflate the effective GP noise wherever a temporal gap of $>3$ days exists. This prevents the GP from artificially shrinking its posterior variance in poorly constrained regions of the light curve. 
We present the result of this interpolation in Supplementary Data Figure \ref{fig:gp}, showing an excellent fit to the data in each band.

\subsection{Spectroscopic reductions}\label{sec:spec_reduction}
We obtained spectra using the facilities and spectrographs listed below:
\begin{itemize}
    \item \textbf{NTT/EFOSC2} -- A first classification spectrum was taken with the ESO Faint Object Spectrograph and Camera 2 (EFOSC2) spectrograph mounted on the 3.58 m New Technology Telescope (NTT) in the ESO observatory at La Silla, Chile. The spectrum was taken during a test run for the Son of X-shooter (SoXS) spectrograph scheduler \citep{Asquini2024a} with the EFOSC2 spectrograph. The spectrum was uploaded to the Transient Name Server\citep{tns} (TNS) and was made public.

    An additional four epochs of EFOSC2 spectra were taken through the European Southern Observatory Spectroscopic Survey of Transient Objects (ePESSTO+) collaboration and were reduced using the standard PESSTO pipeline \citep{Smartt_epessto_2015}.

    \item \textbf{Gemini-S/GMOS} -- We used the Gemini Multi-Object Spectrograph (GMOS) \cite{Hook2004,Gimeno2016} attached to the Gemini South telescope in long-slit mode to observe SN 2024vjm on 2024-09-16 and 2024-09-21, with total exposure times of 1200 s in each epoch. With the B480 grating centred at 540/545 nm, and 2x2 binning, the observations covered 380-750 nm at spectral resolution around 1300. The Gemini \texttt{DRAGONS} software package \cite{Labrie2023} was used to reduce the data following standard procedure, to obtain the reduced spectra calibrated in wavelength and flux.
    
    \item \textbf{FTS/FLOYDS} -- Additional five spectra were obtained by the GSP with the FLOYDS spectrograph mounted on the LCO 2m telescope in the Sliding Springs Observatory in Australia. The spectra were reduced using the LCO \texttt{floydsspec} pipeline, which deals with cosmic ray removal, flagging, spectrum extraction and calibration \citep{FLOYDS}.
    
    \item \textbf{VLT/X-shooter} -- We obtained a single epoch of spectroscopy using the X-shooter wide-band echelle spectrograph \citep{Vernet2011} mounted on the 8.2m Very Large Telescope (VLT) in ESO's Paranal Observatory, Chile. The observations were performed under the program 113.26B1 (PI: Blagorodnova). The SN was obtained near the parallactic angle using a stare mode with slit sizes of 1, 0.9, and 0.6 arcsec for the UBV, VIS, and NIR channels, respectively. 
    The data were reduced using the ESO pipeline \texttt{EsoRex} v.3.13.7 together with the standard X-shooter slit spectrograph pipeline v.3.6.3. Bias frames were taken five days before the observation, and the flat fields five to 15 days before the observation.
    Telluric absorption correction was applied to the VIS arm using a standard template.
    
    As \sn was yet to be classified during the observation, the data were taken to optimize the signal in the optical UBV and VIS bands. Hence, the stare mode was chosen, leading to decreased quality in the NIR band, as telluric emission lines could not be subtracted from adjacent pixels. Additionally, to increase the exposure time, no telluric star observations were taken at a similar airmass to allow telluric subtraction. To mitigate this, since much of the telluric noise in the NIR originates from strong emission lines, we employed a sigma clipping method to subtract strong telluric spikes in the X-shooter NIR spectrum. To achieve decent quality reductions, we used a window of 21 flux entries corresponding to $12.6\,\rm \AA$ bins. We then calculate the mean value of each window and remove any residual flux values that are more than $5 \sigma$ from the mean. We run this process iteratively seven times until no more residuals are found by the algorithm, after which a cleaner spectrum is left for which we can identify SNe features similar to other early Type Iax NIR spectra (see \extmat \ref{sec:spectroscopy}).
    
    \item \textbf{SALT/RSS} -- \sn spectra, obtained by SALT/RSS were observed with a 1.5$''$ wide long-slit, and the PG0900 grism in two tilt positions. The data was reduced using a custom pipeline, called \texttt{RUSALT}, based on the \texttt{PySALT} package \citep{Crawford_pysalt_2010} and standard \texttt{Pyraf} \citep{pyraf_2012} spectral reduction routines. 
    
    \item \textbf{Magellan/IMACS} --  We obtained one optical spectrum of SN\,2024vjm on 2024-11-06 using the Inamori Magellan Areal Camera and Spectrograph \citep[IMACS;][]{Dressler2011} mounted on the 6.5-m Magellan-Baade telescope under decent conditions ($\sim$ 0.8\farcs). The observations consist of three 600-second exposures with a 300 lines/mm grating, resulting in a spectral resolution $R\sim$1000. The spectra were reduced with {\tt IRAF}, including basic data processing (bias subtraction, flat fielding), cosmic-ray removal, wavelength calibration (using arc lamp frames taken immediately after the target observation), and relative flux calibration with a spectroscopic standard observed the same night as the science object.

    \item \textbf{SOAR/ TripleSpec} -- We obtained three NIR spectra of \sn on 2024-09-28, 2024-10-10, and 2024-10-21 using the TripleSpec spectrograph on the SOAR telescope. All spectra were taken in the cross-dispersed mode through AEON queue observations (NOIRLab Program ID: 2024B-237887;  PI: Aravind P. Ravi) using a 1.1 arcsec longslit. Data were reduced using the TripleSpec-specific modification \citep{Kirkpatrick11,TripleSpec} of the commonly used NIR spectral reduction package, \verb|Spextool| \citep{Cushing04}. Both the science and the adjacent (in airmass) A0V standard targets were observed in an ABBA dithering pattern and the latter were used for telluric correction following the standard procedure as described in \citet{Vacca03}.

    To improve upon the original telluric reduction, we apply the same sigma-clipping method applied to the X-shooter spectrum to the SOAR spectra. This marginally improves the quality of the spectra by removing residual telluric spikes from the spectra.
\end{itemize}

\subsection{MUSE reduction}\label{sec:MUSE_red}
The observations were reduced using the ESO MUSE pipeline \citep{Weilbacher2020} run automatically by ESO and uploaded to the ESO archive. We then post-reduce excess sky emission using the Zurich Atmosphere Purge \citep[\texttt{ZAP};][]{Soto2016} code using its \texttt{Python} implementation.
The original astrometric solution derived by the pipeline was offset, likely due to the lack of point sources in the MUSE field of view. We therefore recalibrated the astrometric solution by pegging \sn's coordinates to its centre of light spaxel in the MUSE data.

Alongside spectra of the nearby star-forming regions (see Section \ref{sec:environment}), we have extracted a spectrum of \sn from the MUSE datacube, presented in Extended Data Figure \ref{fig:spec} alongside the other optical spectra taken for this study.
Additionally, we obtained a spectrum from our single VLT/MUSE observation described in \suppmat \ref{sec:MUSE_red}.

\clearpage

\section{Supplementary Data Items}
\setcounter{figure}{0}
\setcounter{table}{0}

\begin{table}[h!]
    \centering
    \captionsetup{name=\supmattwo Table}
    \footnotesize
    \caption{Log of spectra}
    \label{tab:spectra}
    \begin{tabular}{llllll}
        \toprule
        Telescope/Instrument & Observation time (UTC) & Phase (days) & Grating & Slit (arcsec) & Exposure time (s) \\
        \midrule
        NTT/ EFOSC2 & 2024-09-15T00:40:56.00 & -4.32 & gr13 & --- & 2700 \\
        FTS/ FLOYDS & 2024-09-15T10:59:34.538 & -3.89 & red/blu & 2.0 & 3600.153 \\
        Gemini/ GMOS-S & 2024-09-15T23:40:54.650 & -3.19 & B480 & 1.0 & 300 \\
        VLT/ X-Shooter & 2024-09-17T01:03:12.890 & -2.31 & UVB & 1.0 & 2850 \\
        VLT/ X-Shooter & 2024-09-17T01:03:18.050 & -2.31 & VIS & 0.9 & 2880 \\
        VLT/ X-Shooter & 2024-09-17T01:03:21.1465 & -2.31 & NIR & 0.6 & 3000 \\
        FTS/ FLOYDS & 2024-09-20T09:40:02.792 & 1.05 & red/blu & 2.0 & 3600.145 \\
        FTS/ FLOYDS & 2024-09-21T12:17:05.117 & 2.16 & red/blu & 2.0 & 3600.026 \\
        SALT/ RSS & 2024-09-21T18:13:18.330 & 2.41 & PG0900 & --- & 1133.290 \\
        Gemini/ GMOS-S & 2024-09-21T23:38:26.2 & 2.58 & B480 & 1.0 & 1200 \\
        SALT/ RSS & 2024-09-24T18:06:51.693 & 5.40 & PG0900 & --- & 1533.287 \\
        SOAR/ TripleSpec & 2024-09-28T00:06:35.427 & 8.4 & --- & 1.1 & 266.873 \\
        NTT/ EFOSC2 & 2024-10-03T01:58:23.00 & 13.73 & Gr16 & --- & 2699 \\
        SOAR/ TripleSpec & 2024-10-10T00:25:31.077 & 20.42 & --- & 1.1 & 200. \\
        SALT/ RSS & 2024-10-11T19:07:49.104 & 22.45 & PG0900 & --- & 1945.238 \\
        VLT/ MUSE & 2024-10-13T01:10:42.118 & 23.70 & --- & --- & 2827.889 \\
        NTT/ EFOSC2 & 2024-10-14T02:41:48.00 & 24.76 & Gr16 & --- & 2699 \\
        SALT/ RSS & 2024-10-14T18:17:13.196 & 25.41 & PG0900 & --- & 2000.285 \\
        SALT/ RSS & 2024-10-15T18:19:32.107 & 26.41 & PG0900 & --- & 2000.278 \\
        SALT/ RSS & 2024-10-17T18:31:33.722 & 28.42 & PG0900 & --- & 2000.283 \\
        SOAR/ TripleSpec & 2024-10-22T01:01:26.277 & 32.44 & --- & 1.1 & 200. \\
        FTS/ FLOYDS & 2024-10-29T10:09:31.254 & 40.07 & red/blu & 2.0 & 3000.179 \\
        FTS/ FLOYDS & 2024-11-02T10:54:34.597 & 44.10 & red/blu & 2.0 & 3600.186 \\
        Magellan/ IMACS & 2024-11-06T00:27:45 & 47.67 & Gri-300 & 1.0 & 1800 \\
        NTT/ EFOSC2 & 2024-11-13T00:41:40.00 & 54.68 & Gr13 & --- & 2699 \\
        NTT/ EFOSC2 & 2024-11-29T01:13:41.00 & 70.70 & Gr13 & --- & 1799 \\
        \bottomrule
    \end{tabular}
\end{table}

\begin{table}[ht]
\captionsetup{name=\supmattwo Table}
\centering
\footnotesize
\caption{Measured emission line fluxes for the star-forming region}
\label{tab:host_emission_lines}
\begin{tabular}{l c}
\hline
\hline
Line & Flux ($10^{-16}$ erg s$^{-1}$ cm$^{-2}$) \\
\hline
H$\alpha$ & 19.13 $\pm$ 0.86 \\
H$\beta$  & 5.27 $\pm$ 1.14 \\
{[}O III] $\lambda$5008 & 2.46 $\pm$ 1.09 \\
{[}N II] $\lambda$6583 & 7.16 $\pm$ 0.39 \\
{[}S II] $\lambda$6716 & 4.94 $\pm$ 0.41 \\
{[}S II] $\lambda$6732 & 3.15 $\pm$ 0.37 \\
\hline
\end{tabular}
\end{table}


\begin{figure}[h!]
    \centering
    \captionsetup{name=\supmattwo Figure}
    \includegraphics[width=\linewidth]{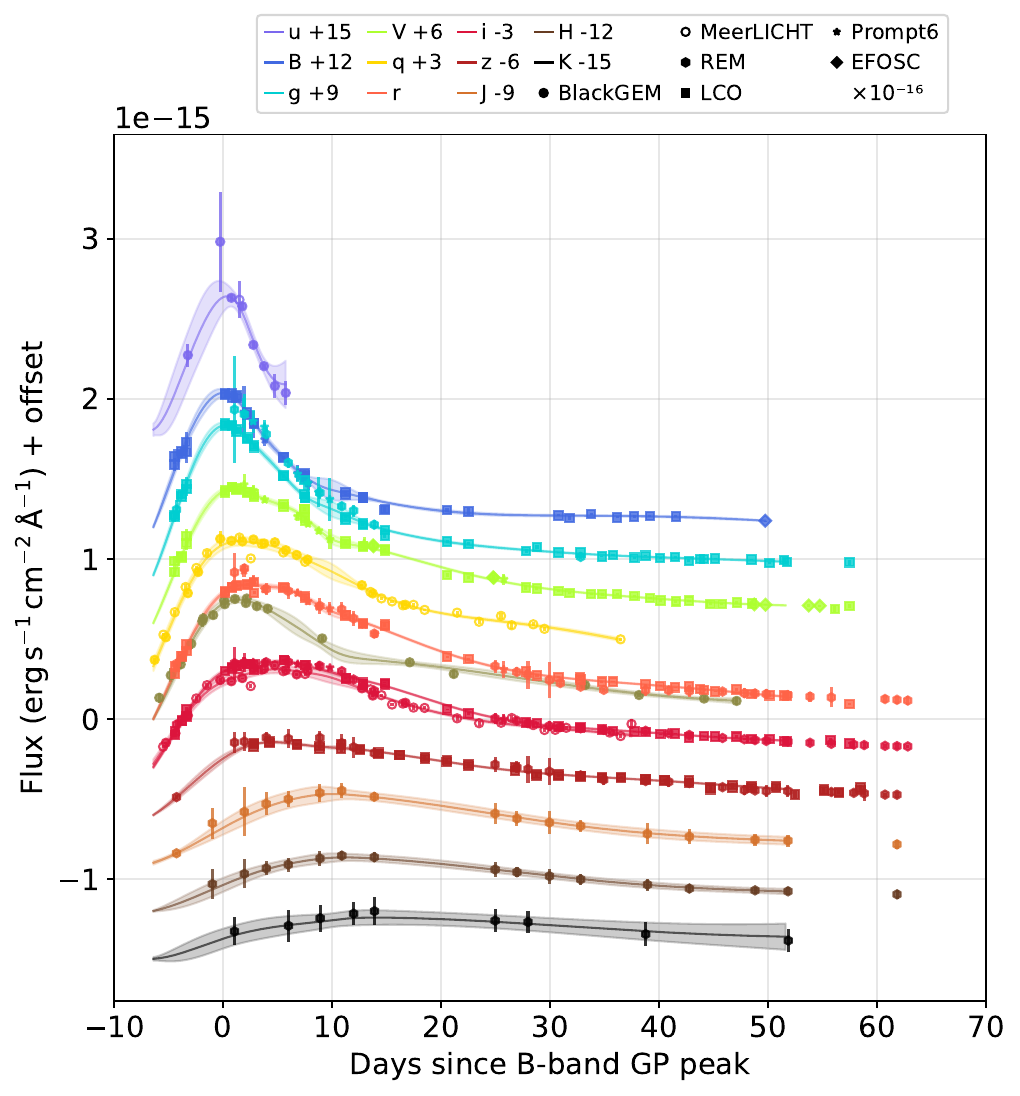}
    \caption{GP interpolation of the SN lightcurve. We find the interpolation (solid lines) to closely follow the data. Calculated $1\sigma$ confidence intervals are presented as transparent bands around the interpolation.}
    \label{fig:gp}
\end{figure}

\clearpage





\bibliography{sn-bibliography}
\end{document}